\documentclass{article}

\usepackage[utf8]{inputenc}
\usepackage[margin=1.5in]{geometry}
\usepackage{amsmath}
\usepackage{amsfonts}
\usepackage{amssymb}
\usepackage{mathrsfs}
\usepackage{subcaption}
\usepackage{xcolor}
\usepackage{soul}
\usepackage{cancel}
\usepackage{amsthm}
\usepackage{mathtools}
\usepackage{natbib}
\usepackage[pdftex]{graphicx}
\usepackage{color}
\usepackage{longtable}
\usepackage{tikz}
\usepackage{comment}
\usepackage{bbm}
\usepackage{tabularx}
\usepackage{booktabs}
\usepackage{multirow}
\usepackage[normalem]{ulem}
\usepackage{url}

\definecolor{codegreen}{rgb}{0,0.6,0}
\definecolor{codegray}{rgb}{0.5,0.5,0.5}
\definecolor{codepurple}{rgb}{0.58,0,0.82}
\definecolor{backcolour}{rgb}{0.95,0.95,0.92}

\newcommand{\be}{\begin{equation}}
\newcommand{\ee}{\end{equation}}
\def \bsp {\begin{split}}
\def \esp {\end{split}}
\def \bea {\begin{eqnarray}}
\def \eea {\end{eqnarray}}

\title{A Unified Framework for Fast Large-Scale Portfolio Optimization}

\author{\vspace{-5mm}{Weichuan Deng}$^{a}$ \hspace*{2mm}  %\addtocounter{footnote}{-2}
{Pawe\l \ Polak}$^{a, c}$\footnote{Corresponding author at Department of Applied Mathematics and Statistics at Stony Brook University, United States. E-mail address: \texttt{pawel.polak@stonybrook.edu}.}\hspace*{2mm}
{Abolfazl Safikhani}$^{b}$\hspace*{2mm}
{Ronakdilip Shah}$^{a}$
\\[5mm]
$^{a}$\hspace{-2mm}\textit{\vspace{-2mm} \small Department of Applied Mathematics and Statistics, Stony Brook University, United States}\\[2mm]
$^{b}$\hspace{-2mm}\textit{\vspace{-2mm} \small Department of Statistics, George Mason University, United States}\\[2mm]
$^{c}$\hspace{-2mm}\textit{\vspace{-2mm} \small Institute for Advanced Computational Science, Stony Brook University, United States}}

\begin{document}
\date{\today }
\maketitle
\begin{abstract}
\noindent{We introduce a unified framework for rapid, large-scale portfolio optimization that incorporates both shrinkage and regularization techniques. This framework addresses multiple objectives, including minimum variance, mean-variance, and the maximum Sharpe ratio, and also adapts to various portfolio weight constraints. For each optimization scenario, we detail the translation into the corresponding quadratic programming (QP) problem and then integrate these solutions into a new open-source Python library. Using 50 years of return data from US mid to large-sized companies, and 33 distinct firm-specific characteristics, we utilize our framework to assess the out-of-sample monthly rebalanced portfolio performance of widely-adopted covariance matrix estimators and factor models, examining both daily and monthly returns. These estimators include the sample covariance matrix, linear and nonlinear shrinkage estimators, and factor portfolios based on Asset Pricing (AP) Trees, Principal Component Analysis (PCA), Risk Premium PCA (RP-PCA), and Instrumented PCA (IPCA). Our findings emphasize that AP-Trees and PCA-based factor models consistently outperform all other approaches in out-of-sample portfolio performance. Finally, we develop new $\ell_1$ and $\ell_2^2$ regularizations of factor portfolio norms which not only elevate the portfolio performance of AP-Trees and PCA-based factor models but they have a potential to reduce an excessive turnover and transaction costs often associated with these models.}
\end{abstract}

\vspace{1cm}

\newpage

\section{Introduction}\label{sec:introduction}
Institutional investors often manage portfolios comprising hundreds of assets, and the performance of such portfolios is evaluated through frequent backtesting exercises. These backtests rely different models and numerous optimizations, performed repetitively using a rolling-window scheme and a long history of return data. In this paper, we introduce a unified framework for portfolio optimization. This framework employs Quadratic Programming (QP) methods to calculate portfolios with $\ell_1$ and $\ell_2^2$ regularization, long-short constraints, and various portfolio objective functions such as minimum-variance, mean-variance, and maximum-Sharpe ratio. Owing to the efficiency of the QP optimization algorithms, our proposed models are suitable for the realistic settings of large-dimensional portfolios. These can be applied repeatedly in a rolling window scheme, facilitating backtesting evaluations and refining investment strategies.

Our portfolio optimization framework requires the estimation of a mean vector and a covariance matrix. The two main approaches for the latter are shrinkage covariance matrix estimation and financial factors modeling. The former uses information contained in the assets returns only. It has been studied extensively starting from linear shrinkage covariance matrix estimator by \cite{Ledoit:04}, nonlinear shrinkage estimators such as \cite{Ledoit:12}, and \cite{LedoitWolf:20}, up to the most recent nonlinear quadratic shrinkage estimator proposed by \cite{LedoitWolf:22} (see Section \ref{sec:covariance_model} for more details). The latter approach uses common risk factors with financial or economic interpretations,
which are well-known to capture large amounts of variation in the returns.  Among the most famous models are CAPM-model of  \cite{treynor1961market}, \cite{sharpe1964capital}, \cite{lintner1965security}, and \cite{mossin1966equilibrium}, the three-factor, four-factor, and the five-factor model by \cite{FAMA1993}, \cite{carhart1997persistence} and \cite{fama2015five}, respectively. The extensions of these models under the non-Gaussianity assumption for the asset returns and factors are given in \citet{hediger2021heterogeneous}. There is also the relative momentum factor, which extends the three-factor model. It was first introduced and analyzed by \cite{jegadeesh1993mom}, see also \citet{FracMom:22} and the references therein for momentum-based portfolio strategy without crashes.

While the aforementioned classical common risk factors remain among the most important, a large literature now exists on determining the inclusion of particular factors from the dozens, if not hundreds, available: see, e.g., 
\cite{bai2002determining}, \cite{stock2002forecasting}, \cite{tsai2010constrained}, \cite{bai2013principal}, \cite{bai2016efficient}, and the references therein. The amount of available alternative data, coupled with advancements in computational power and statistical techniques, such as the estimation of sparse models as in \citet{tibshirani1996} and \citet{Hastie:2015}, has led to the proliferation of different factor models, giving rise to what \citet{feng2020taming} describes as a  ``zoo of factors.''

In this paper, we consider a large universe of liquid US stocks and 33 asset-specific characteristics, as listed in Table \ref{table:characteristics} in the Appendix. To extract relevant information from this large number of factors while capturing the dynamics in the dependency between factors and returns in a large portfolio of assets, we use different models, such as: Asset Pricing (AP) Trees introduced in \citet{bryzgalova2020forest}, and three different Principal Component Analysis (PCA) based factor models that invest in leading factor portfolios including the PCA on the factor portfolios, the Risk Premium PCA (RP-PCA) introduced in \citet{lettau2020factors}, and the Instrumented PCA (IPCA) from \citet{KELLY2019501}. All of these papers show that their asset-specific factor based models outperform the common risk factors models mentioned earlier in terms of higher in-sample and predicted $R^2$ values, leading to higher out-of-sample portfolio performance. Recently, \citet{goyal2022equity} used IPCA to explain the returns of option contracts and achieved a significantly better out-of-sample $R^2$. Motivated by these successes of these recent factor-based models and their flexibility in capturing information from a large number of stock-specific characteristics, we forgo the aforementioned common risk factors models and focus on the AP-Trees and PCA-based models in our unified portfolio optimization framework. We compare these emerging models and the aforementioned shrinkage approaches in portfolio optimization with liquid stocks under realistic portfolio constraints.

In \citet{lettau2020factors} and \citet{KELLY2019501}, the portfolio performance of PCA-based models is evaluated using the tangent portfolio. This is a closed-form portfolio that permits unbounded long and short positions in individual assets, as well as highly leveraged long-short portfolio strategies. In this paper, we contrast the portfolio performance of the PCA-based models with commonly used benchmarks, such as the shrinkage covariance matrix estimator. We employ a rolling window exercise on an extensive history of a large set of liquid US equity returns, excluding small and micro-caps. We also apply realistic constraints on individual positions and long-short strategies to prevent highly concentrated positions and excessively leveraged portfolios. Our portfolio performance largely agrees with the original results in \citet{lettau2020factors} and \citet{KELLY2019501}. But this more grounded setup further illustrates the versatility of the proposed unified portfolio optimization framework, making it relevant to the practical portfolio challenges faced by large institutional investors.

Our paper presents four primary contributions. First, we introduce a unified framework for large-scale, rapid portfolio optimization that incorporates realistic constraints and innovative regularizations to enhance investment performance. This framework is particularly relevant for institutional investors managing portfolios with hundreds or even thousands of assets, facilitating cost-efficient investment decisions. As a practical tool, we've made our Python implementation of this framework available as open-source code online.\footnote{The latest version of the code can be found at \url{https://github.com/PawPol/PyPortOpt}} Second, we offer fresh insights into the performance of the recently discussed AP-Trees and PCA-based models. Third, our framework supports a multitude of portfolio problem combinations, varying in portfolio objective functions, regularizations, and constraints. This includes the $\ell_1$ and $\ell_2^2$ regularized portfolio problems, as introduced by \citet{demiguel2009generalized} for the minimum-variance portfolio. We expand upon this by introducing the $\ell_1$+$\ell_2^2$ regularized maximum-Sharpe ratio portfolios and the comprehensive $\ell_1$+$\ell_2^2$ regularized mean-variance portfolio frontier. Lastly, within the scope of AP-Trees and PCA-based models, we demonstrate how to apply our novel regularizations to both managed portfolios and individual stocks. We further illustrate how these new regularizations result in superior performance, leading to more stable and streamlined portfolio positions. Importantly, we show how to solve all of these optimization problems using QP methods.

The rest of the paper is structured as follows. Section \ref{sec:portfolio} presents our comprehensive framework for portfolio optimization. Section \ref{sec:covariance_model} elaborates on the various covariance matrix estimators discussed in this study. Section \ref{sec:maxSharpeRatioPortfolio_PCAbased} introduces a novel regularization for factor-based portfolio optimization challenges with an emphasizes on maximu Sharpe ratio portfolio. Empirical comparisons of the estimators and models across distinct portfolio optimization problems are detailed in Section \ref{sec:empirics}. Concluding observations are given in Section \ref{sec:conclusions}. The Appendix provides details on the asset-specific factors.

\section{Portfolio Optimization Framework}\label{sec:portfolio}
We consider a universe of $N$ assets, with prices observed over a given period of time with $T$ observations. Let $P_{t,i}$ be the price of asset $i=1,\ldots,N$ at time index $t=1,\ldots,T$, where the time index $t$ corresponds to a fixed unit of time such as days, weeks, or months. The corresponding simple returns\footnote{In the empirical analysis we work with dividend and split adjusted simple returns.} (also known as linear or net returns) are given by $R_{t,i} =\frac{P_{t,i} - P_{t-1,i}}{P_{t-1,i}}=\frac{P_{t,i}}{P_{t-1,i}}-1$, and the log-returns (also known as continuously compounded returns) are 
$r_{t,i} = \log \frac{P_{t,i}}{P_{t-1,i}} = \log(1+R_{t,i}).$

We denote the vector of $\log$-returns of $N$ assets at time $t$ with $\mathbf{r}_t\in \mathbb{R}^{N}$. It is a multivariate stochastic process with conditional mean and covariance matrix denoted by
\[ \mathbb{E} [\mathbf{r}_t\mid \mathcal{F}_{t-1}]=\boldsymbol{\mu}_t =\begin{bmatrix}\mu_{t,1}\\ \vdots\\ \mu_{t,N}\end{bmatrix}\nonumber \]
and
\[ Cov[\mathbf{r}_t\mid \mathcal{F}_{t-1}] =  \mathbb{E}
[(\mathbf{r}_t - \boldsymbol{\mu}_t)(\mathbf{r}_t - \boldsymbol{\mu}_t)^T \mid \mathcal{F}_{t-1}]=\boldsymbol{\Sigma}_t =\begin{bmatrix}\sigma_{t,11} &\cdots& \sigma_{t,1N}\\ \vdots&\ddots&\vdots\\
\sigma_{t,N1}&\cdots&\sigma_{t,NN}\end{bmatrix}\nonumber, \]
where $\mathcal{F}_{t-1}$ denotes the previous historical data. In this work, except for the IPCA model, we will drop the subscript $t$ on the mean and covariance matrix since all models assume $iid$ returns. For more general multivariate time-series models of returns with the dynamics in the conditional mean and covariance matrix together with their applications in portfolio optimization, we refer to \citet{PaPo:15c}, \citet{paolella2019regime}, and \citet{Paolella:2021}.

The investment portfolio is usually summarized by an $N$-vector of weights $\mathbf{w} = [w_1,\ldots,w_N]^{\prime}$ indicating the fraction of the total wealth of the investor held in each asset. If the investor is assumed to hold her total wealth in the portfolio, then $\mathbf{w}'\mathbf{1}_N = 1$, where $\mathbf{1}_N$ denotes an $N$-vector of ones. The corresponding portfolio return $r_t(\mathbf{w}) = \mathbf{w}^{\prime}\mathbf{r}_t$ is a random variable with the mean and variance given by $\mu_{\mathbf{w}} = \mathbb{E}[r_t(\mathbf{w})] = \mathbf{w}^{\prime}\boldsymbol{\mu}$ and $\sigma^2_{\mathbf{w}} = Var[r_t(\mathbf{w})] = \mathbf{w}^{\prime}\boldsymbol{\Sigma}\mathbf{w}$, respectively.

The general theory of portfolio optimization, as introduced in a seminal work by \citet{Ma52}, summarizes the trade-off between risk and investment return using the portfolio's mean and variance. In particular, for a given choice of target mean return \(\alpha_0\), in Markowitz portfolio optimization, one chooses the optimal portfolio as 
\begin{equation}\label{eq:meanVariancePortOpt}
\mathbf{w}^* = \arg\min_{\mathbf{w}\in\mathcal{W}}\frac{1}{2}\mathbf{w}^\prime\mathbf{\Sigma}\mathbf{w},
\end{equation}
where  $\mathcal{W}:=\left\{\mathbf{w}\in \mathbb{R}^N:\mathbf{w}^{\prime}\boldsymbol{\mu} \geq \alpha_0\mbox{ and } \mathbf{w}^{\prime}\mathbf{1}_{N} = 1\right\}$ is a set of constraints on the portfolio weights which correspond to a fully invested portfolio with the expected return above the $\alpha_0$ threshold. Under these constraints, \eqref{eq:meanVariancePortOpt} has a closed-form solution given by
\begin{equation}\label{eq:long_short_closed_form}
\mathbf{w}^* = \left\{{B}\boldsymbol{\Sigma}^{-1}\mathbf{1} - {A} \boldsymbol{\Sigma}^{-1}\boldsymbol{\mu} + \alpha_0({C}\boldsymbol{\Sigma}^{-1}\boldsymbol{\mu} - {A}\boldsymbol{\Sigma}^{-1} \mathbf{1})\right\}/{D},
\end{equation}
where ${A}=\boldsymbol{\mu}\boldsymbol{\Sigma}^{-1}\mathbf{1} =\mathbf{1}^{\prime} \boldsymbol{\Sigma}^{-1}\boldsymbol{\mu}$, ${B}=\boldsymbol{\mu}^{\prime}\boldsymbol{\Sigma}^{-1}\boldsymbol{\mu}$, ${C}=\mathbf{1}^{\prime}\boldsymbol{\Sigma}^{-1}\mathbf{1} $, ${D}={B}{C}-{A}^2$.

\begin{figure}
\centering
\subfloat{
\resizebox*{15.5cm}{!}{\includegraphics{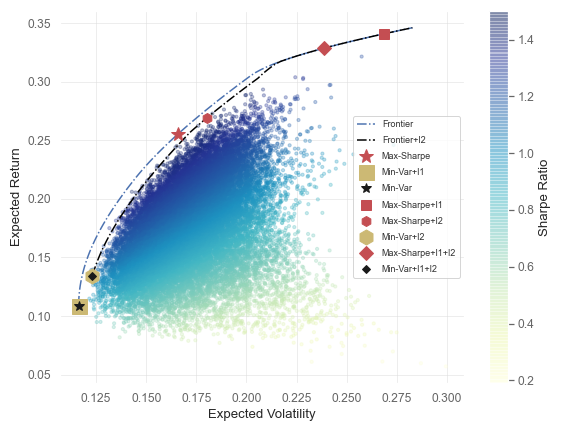}}}\hspace{5pt}
\caption{ \protect \footnotesize Portfolio frontier (with and without $\ell_2^2$ regularization) and all of the optimal portfolios considered in our portfolio framework with the long-only constraints and $\ell_1$, $\ell_2^2$, and $\ell_1$+$\ell_2^2$ regularization for eight stocks (AMZN, MSFT, GOOGL, F, TM, AAPL, KO, and PEP), with the mean and covariance matrix estimated using daily returns over eight years (2015/01/01-2022/01/01). Among them are two optimal portfolios: the minimum-variance portfolio and the maximum Sharpe ratio portfolio, and a collection of random portfolios.} \label{fig:portfoliofrontier}
\end{figure}

The minimum-variance portfolio ($Min$-$Var$ in Figure \ref{fig:portfoliofrontier}) is a solution to \eqref{eq:meanVariancePortOpt} with 
$\mathcal{W}:=\left\{\mathbf{w}\in \mathbb{R}^N: \mathbf{w}^{\prime}\mathbf{1}_{N} = 1\right\}$. The solution to this problem also has a closed-form expression given by
\begin{equation}\label{eq:minimumvariance_closed_form}
    \mathbf{w}^* = \boldsymbol{\Sigma}^{-1} \mathbf{1}/C,
\end{equation}
where $C$ is defined above. However, when short-selling is not allowed, i.e., $\mathbf{w}\geq \boldsymbol{0}_N$, or when it is constrained, e.g., as in Section \ref{sec:long_short_portfolio}, then the optimization problem \eqref{eq:meanVariancePortOpt} does not have a closed-form solution and needs to be solved numerically.

Nevertheless, \eqref{eq:meanVariancePortOpt} is a QP problem with convex constraints (hence also a convex problem). It has closed-form expressions for the gradient and hessian of the objective function, and a unique global optimal portfolio satisfying the constraints in $\mathcal{W}$. In particular, by changing $\alpha_0$, one can derive a whole portfolio frontier of optimal investments $\mathbf{w}^*(\alpha_0)$ summarizing the risk-return trade-off. 

Following \citet{li2015sparse}, we can reinterpret the mean-variance portfolio optimization problem as a linear regression with \(N\) independent variables and \(N\) observations. This relationship can be expressed as:
\begin{equation}\label{eq:linear_represented}
    \mathbf{y} = \mathbf{Xw} + \mathbf{e},
\end{equation}
where \(\mathbf{y} = \frac{1}{\sqrt{\gamma}}\boldsymbol{\Sigma}^{-\frac{1}{2}}\boldsymbol{\mu}\), \(\mathbf{X} = \sqrt{\gamma}\boldsymbol{\Sigma}^{\frac{1}{2}}\), \(\mathbf{e}\) represents a vector of random errors, and \(\gamma > 0\) is the risk aversion coefficient (Lagrange multiplier) associated with the \(\alpha_0\) threshold in \(\mathcal{W}\) described above. The least squares estimator of \(\mathbf{w}\), given by \(\hat{\mathbf{w}}_{OLS} = (\mathbf{X}^T\mathbf{X})^{-1}(\mathbf{X}^T\mathbf{y})\), corresponds to the closed-form optimal portfolio weight when the constraint \(\mathbf{w}^{\prime}\mathbf{1}_{N} = 1\) is omitted. In other words, \(\hat{\mathbf{w}} = \frac{1}{\gamma}\boldsymbol{\Sigma}^{-1}\boldsymbol{\mu}\). In practice, $\boldsymbol{\Sigma}$ and $\boldsymbol{\mu}$ are unknown and they are replaced by their (random) estimators. Thus, the principles of linear regression can be naturally extended to portfolio optimization. In a similar vein, the theories of \(\ell_1\) and \(\ell_2^2\) regularized regression can be directly related to the regularized portfolio optimization problem. When the portfolio constraint \(\mathbf{w}^{\prime}\mathbf{1}_{N} = 1\) is incorporated, this mirrors the analogous constraint in the least squares problem.

Figure \ref{fig:portfoliofrontier} presents two long-only mean-variance portfolio efficient frontiers, both with and without the $\ell_2^2$ regularization discussed in Section \ref{sec:l2_constraint_portfolio_norms}. For varying levels of portfolio variances, the expected return of the top-performing portfolio is plotted. Alongside these frontiers, we illustrate various optimal portfolios discussed in this paper. Additionally, a cloud of points represents the means and variances of 25,000 randomly drawn $iid$ Dirichlet distributed portfolios. Specifically, each portfolio weight vector $\mathbf{w}_k$ is independently and identically distributed as $Dir(\mathbf{1}_N)$ for $k=1,\ldots,25000$. In this example, the portfolios are comprised of eight stocks from the US market with tickers: AMZN, MSFT, GOOGL, F, TM, AAPL, KO, and PEP. The mean and covariance matrix are estimated using daily returns spanning the period from 2015-01-01 to 2022-01-01. Such a low dimensional portfolio problem is common in the aforementioned PCA-based models which invest into $K$ factor portfolios that are mapped into the individual assets.

\begin{figure}
\centering
\subfloat[N = 10\label{fig:1a}]{%
\resizebox*{6.5cm}{!}{\includegraphics{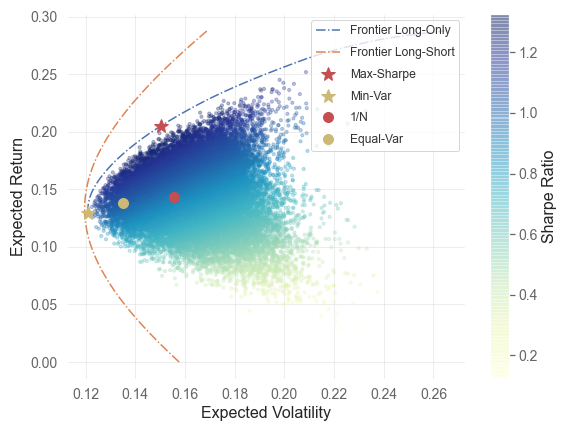}}}\hspace{5pt}
\subfloat[N = 20\label{fig:1b}]{%
\resizebox*{6.5cm}{!}{\includegraphics{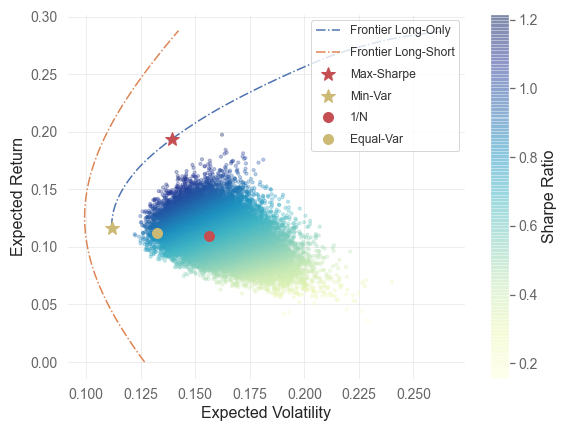}}}\hspace{5pt}
\subfloat[N = 50\label{fig:1c}]{%
\resizebox*{6.5cm}{!}{\includegraphics{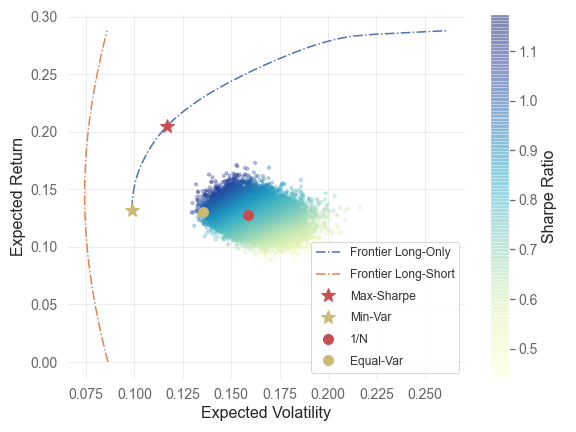}}}\hspace{5pt}
\subfloat[N = 500\label{fig:1d}]{%
\resizebox*{6.5cm}{!}{\includegraphics{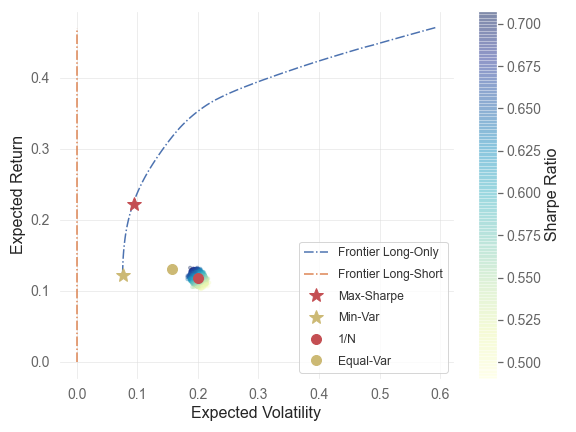}}}\hspace{5pt}
\caption{ \protect \footnotesize Four plots of two different portfolio frontiers (long-only and the closed-form long-short from \eqref{eq:long_short_closed_form}), together with different optimal long-only portfolios (maximum-Sharpe ratio \eqref{eq:maximumSharperatio} and minimum-variance), the equally weighted portfolio ($1/N$), equal volatility contribution portfolio (Equal-Var), and $25000$ $iid$ Dirichlet distributed portfolios $\mathbf{w}\sim Dir(\mathbf{1}_N)$, for different number of assets $N=10, 20, 50, 500$ selected from the largest market-capitalization stocks in the US market. Mean and covariance matrix estimated from 10 years of daily returns (2520 observations).} \label{fig:differentassetsportfolio}
\end{figure}

In practice, it is often the case that the investment portfolio consist of a much larger number of assets than in the example above. 
Figure \ref{fig:differentassetsportfolio} illustrates portfolio frontiers together with $25000$ $iid$ Dirichlet distributed\footnote{Here we use Dirichlet distributed random vectors to guarantee uniform sampling on the $N$ dimensional simplex ($\mathbf{w}'\mathbf{1}_N=1$). The results for the weights sampled from uniform distribution normalized on the simplex, i.e., $\mathbf{w}= \mathbf{x}/(\mathbf{x}'\mathbf{1}_N)$, where $\mathbf{x}=[x_1,\ldots,x_N]$ and $x_i\overset{iid}{\sim}  U([0,1])$; and for the weights sampled from the absolute value of standard normal distribution normalized on the simplex, i.e., $\mathbf{w}= \left|\mathbf{x}\right|/\left\|\mathbf{x}\right\|_1$, where $\mathbf{x}=[x_1,\ldots,x_N]$ and $x_i\overset{iid}{\sim}  N(0,1)$ are similar.} portfolios $\mathbf{w}_k\overset{iid}{\sim} Dir(\mathbf{1}_N)$, for $k=1,\ldots,25000$, and for the different number of assets $N=10, 20, 50, 500$ selected from the largest market-capitalization stocks in the US market with mean and covariance matrix estimated using ten years of daily returns (which is a much larger number of observations than all of our monthly data used in Section \ref{sec:empirics}). As can be seen from different panels in Figure \ref{fig:differentassetsportfolio}, the dimensionality of the portfolio has two major impacts. First, the larger the assets universe, the further away and the more concentrated around $1/N$ are the random portfolios. This shows that in a large-dimensional setup without proper portfolio optimization, one cannot expect to achieve any optimal risk-reward profile and that even the $1/N$ portfolio, which is so often advocated as a naive-diversification and well-performing portfolio---see \citealp{DGU:09} and the references therein---is in fact as good as any random guess. Figure \ref{fig:differentassetsportfolio} also depicts the equal volatility contribution portfolio, which is a special case of the risk parity portfolio (see, e.g., \citealp{roncalli_RP:13} and \citealp{PaPoPoWa:19}). It is slightly better than random portfolios or the $1/N$. However,  based on the distance between the equal volatility contribution portfolio and the mean-variance portfolio frontier, and even with the uncertainty about the actual frontier, there is still a lot of opportunity for improved portfolio allocation. Second, the closed-form long-short frontier from equation \eqref{eq:long_short_closed_form}, depicted with dotted black lines in all the panels of Figure~\ref{fig:differentassetsportfolio}, is becoming almost vertical compared with the long-only portfolio when the number of assets increases. Therefore, small changes in the optimal portfolio volatility translate to theoretically disproportionately large gains in the expected returns of the optimal portfolio. This implies that estimates of the optimal portfolio weights are sensitive to new data points, and the weights can change a lot over the consecutive rolling windows. This is the artifact of high-dimensionality and relatively close to non-singular covariance matrix estimates. Proper covariance matrix estimation in high dimensions and long-short constraints help in avoiding these over-leveraged and unrealistic but theoretically optimal portfolios.

Figure \ref{fig:differentassetsportfolio} depicts portfolio frontiers alongside 25,000 $iid$ Dirichlet distributed\footnote{We utilize Dirichlet distributed random vectors to ensure uniform sampling on the $N$ dimensional simplex ($\mathbf{w}'\mathbf{1}_N=1$). The results from weights sampled from the uniform distribution normalized on the simplex (i.e., $\mathbf{w}= \mathbf{x}/(\mathbf{x}'\mathbf{1}_N)$, where $\mathbf{x}=[x_1,\ldots,x_N]$ and $x_i\overset{iid}{\sim}  U([0,1]$); and from the weights derived from the absolute value of the standard normal distribution normalized on the simplex (i.e., $\mathbf{w}= \left|\mathbf{x}\right|/\left\|\mathbf{x}\right\|_1$, where $\mathbf{x}=[x_1,\ldots,x_N]$ and $x_i\overset{iid}{\sim}  N(0,1)$) align closely.} portfolios $\mathbf{w}_k\overset{iid}{\sim} Dir(\mathbf{1}_N)$, for $k=1,\ldots,25000$. The assets number varies as $N=10, 20, 50, 500$, chosen from the largest market-capitalization stocks in the US market. The mean and covariance matrix are derived from ten years of daily returns, a period significantly longer than our monthly data in Section \ref{sec:empirics}.

From the varying panels in Figure \ref{fig:differentassetsportfolio}, we discern two significant implications of portfolio dimensionality. First, as the assets universe expands, random portfolios veer further from and concentrate more around the $1/N$ mark. This suggests that without appropriate portfolio optimization in high-dimensional setups, achieving any optimal risk-reward profile is challenging. Even the frequently endorsed $1/N$ portfolio, often hailed for naive-diversification and robust performance (see \citealp{DGU:09} and the cited references), performs equivalently to a random guess. The figure also presents the equal volatility contribution portfolio, a variant of the risk parity portfolio (\citealp{roncalli_RP:13} and \citealp{PaPoPoWa:19}). While it slightly outperforms random portfolios and the $1/N$, the gap between this portfolio and the mean-variance portfolio frontier indicates a lot of room for improved portfolio allocation.

Secondly, the closed-form long-short frontier, represented by \eqref{eq:long_short_closed_form} and illustrated with dotted black lines in all the panels in Figure~\ref{fig:differentassetsportfolio}, appears almost vertical in relation to the long-only portfolio as assets increase. Consequently, marginal shifts in optimal portfolio volatility can lead to theoretically substantial hikes in the expected returns of the optimal portfolio. This highlights the sensitivity of optimal portfolio weight estimates to new data points, with weights potentially exhibiting significant variations across consecutive rolling windows. Such behavior stems from high dimensionality and proximate non-singular covariance matrix estimates. Effective covariance matrix estimation in expansive dimensions, combined with long-short constraints, counters these over-leveraged yet theoretically optimal portfolios.

In practice the true mean vector and covariance matrix are unknown, and one needs to rely on their estimates. Financial markets, especially at low frequencies, are highly efficient---or, as suggested by \citet{Pedersen:15}, they are ``efficiently-inefficient''. 
We do not attempt to construct individual stocks prediction signals---for that we refer to recent results in \citet{FracMom:22}. Instead, we focus on various mean and covariance matrix shrinkage estimators as well as different factor portfolios. The former address the bias-variance trade-off, aiming to construct biased estimators that minimize the mean-square error and perform better out-of-sample. The latter  offers conditional predictions of expected returns based on asset characteristics. As we will demonstrate, the factor portfolios significantly enhance the signal-to-noise ratio, leading to more accurate mean predictions and higher out-of-sample performance. However, before we turn to stock returns models, we introduce the rest of our general portfolio optimization framework.

\subsection{Portfolio Constraints}\label{sec:minimumvarianceportfolio}

The set of feasible portfolio weights $\mathcal{W}:=\left\{\mathbf{w}\in \mathbb{R}^N: \mathbf{w}^{\prime}\mathbf{1}_{N} = 1\right\}$ usually includes additional constraints. Among the most commonly used are:
\begin{itemize}
    \item Long only: 
    $$\mathcal{W}:=\left\{\mathbf{w}\in \mathbb{R}^N: \mathbf{w}^{\prime}\mathbf{1}_{N} = 1 \mbox{ and } w_i\geq 0, \forall i\right\}.$$
    \item Asset specific holding constraints: 
    $$\mathcal{W}:=\left\{\mathbf{w}\in \mathbb{R}^N: \mathbf{w}^{\prime}\mathbf{1}_{N} = 1  \mbox{ and } L_i\leq w_i\leq U_i, \forall i\right\},$$
    where $\mathbf{U}=(U_1,\ldots,U_N)$ and $\mathbf{L}=(L_1,\ldots,L_N)$ are upper and lower bounds for the $N$ portfolio positions.
    \item Turnover constraints: 
    \begin{itemize}
        \item[-] for individual assets limits
        $$\mathcal{W}:=\left\{\mathbf{w}\in \mathbb{R}^N: \mathbf{w}^{\prime}\mathbf{1}_{N} = 1  \mbox{ and } |\Delta w_i|\leq U_i, \forall i\right\},$$
        where $\Delta w_i$ denotes the change in the portfolio weight from the current position to the optimal value and $U_i$ are the turnover limits for individual positions;
    \item[-] for the total portfolio limit
    $$\mathcal{W}:=\left\{\mathbf{w}\in \mathbb{R}^N: \mathbf{w}^{\prime}\mathbf{1}_{N} = 1  \mbox{ and } \left\|\Delta\mathbf{w}\right\|_1 = \sum_{i=1}^N|\Delta w_i|\leq U_*\right\},$$
    where $U_*$ is the turnover limit for the entire portfolio.
    \end{itemize}
    \item Benchmark exposure constraints: 
     $$\mathcal{W}:=\left\{\mathbf{w}\in \mathbb{R}^N: \mathbf{w}^{\prime}\mathbf{1}_{N} = 1 \mbox{ and }\left\|\mathbf{w} - \mathbf{w}_{B}\right\|_1 = \sum_{i=1}^N|w_i - w_{B,i}|\leq U_B\right\},$$
    where, $\mathbf{w}_B$ are the weights of the benchmark portfolio, and $U_B$ is the total error bound.
    \item Tracking error constraints: for a given benchmark portfolio $B$ with  weights $\mathbf{w}_B$, 
    $r_B = \mathbf{w}_B' \mathbf{r}$ is the return of the benchmark portfolio, e.g., S\&P 500 Index, NASDAQ 100, Russell 1000/2000. One can compute the variance of the Tracking Error $Var(TE) = (\mathbf{w}-\mathbf{w}_B)'\boldsymbol{\Sigma}(\mathbf{w}-\mathbf{w}_B),$ and include the corresponding constraint into to the set of feasible portfolio weights 
     $$\mathcal{W}:=\left\{\mathbf{w}\in \mathbb{R}^N: \mathbf{w}^{\prime}\mathbf{1}_{N} = 1 \mbox{ and }(\mathbf{w}-\mathbf{w}_B)'\boldsymbol{\Sigma}(\mathbf{w}-\mathbf{w}_B)\leq \sigma_{TE}^2\right\},$$
     where $\sigma_{TE}^2>0$ is the variance tracking-error of the portfolio.
    \item Risk factor constraints: estimate the risk factors exposure for all the assets in the portfolio, e.g., via the following regression (see \eqref{eq:Fama_French_Factor_Model} for details)
    \[r_{i,t} = \alpha_i + \sum_{k=1}^K \beta_{i,k} f_{k,t} + \epsilon_{i,t}.\] 
    Given these estimates, one can 
    \begin{itemize}
    \item[(i)] constrain the exposure to a given factor $k$ by 
    $$\mathcal{W}:=\left\{\mathbf{w}\in \mathbb{R}^N: \mathbf{w}^{\prime}\mathbf{1}_{N} = 1 \mbox{ and }|\sum_{i=1}^N \beta_{i,k}w_i|\leq U_k\right\}.$$
    \item[(ii)] neutralize the exposure to all the risk factors by  
    $$\mathcal{W}:=\left\{\mathbf{w}\in \mathbb{R}^N: \mathbf{w}^{\prime}\mathbf{1}_{N} = 1 \mbox{ and } |\sum_{i=1}^N \beta_{i,k}w_i|=0\ \forall k\right\}.$$
    \end{itemize}
\end{itemize}
All the constraints listed above (including those that involve the absolute value function---see the remarks in Section \ref{sec:l1_constraint_portfolio_norms})  can be written as linear or quadratic constraints, i.e.,
\begin{itemize}
        \item linear constraints: we can specify $N$-columns matrices $A_w$ and $A_B$  and vectors $u_w$, $u_B$ to introduce linear inequality constraints for the relative positions between the assets or the benchmark
     $$\mathcal{W}:=\left\{\mathbf{w}\in \mathbb{R}^N: \mathbf{w}^{\prime}\mathbf{1}_{N} = 1 \mbox{ and } A_w\mathbf{w}\leq u_w,\ \ A_B(\mathbf{w}-\mathbf{w}_B)\leq u_B\right\}.$$
    \item quadratic constraints: we can specify $N\times N$ matrices $Q_w$, $Q_B$ and scalars $q_w$, $q_B$ to build constraints
    $$\mathcal{W}:=\left\{\mathbf{w}\in \mathbb{R}^N: \mathbf{w}^{\prime}\mathbf{1}_{N} = 1 \mbox{ and } \mathbf{w}'Q_w\mathbf{w}\leq q_w,  (\mathbf{w}-\mathbf{w}_B)'Q_B(\mathbf{w} - \mathbf{w}_B) \leq q_B\right\}.$$
\end{itemize}
Once the constraints are converted into these standard forms, they can be easily combined and incorporated into our portfolio optimization framework. We consider next, a different type of constraint that is often incorporated into portfolio optimization using the method of Lagrange multipliers. These constraints are not imposed by the portfolio manager because of her trading goals or position requirements. They are added because they are a form of regularization of the problem in high dimensions, and they help to improve the out-of-sample portfolio performance in large dimensions.

\subsection{Portfolio Optimization with $\ell_2^2$ Penalized Portfolio Norms}\label{sec:l2_constraint_portfolio_norms}
Consider now an $\ell_2^2$-constrained (also called the ridge penalty) portfolio optimization problem for the minimum-variance portfolio \eqref{eq:meanVariancePortOpt}. Using the method of Lagrange multipliers, we can write the corresponding optimization problem as
\begin{equation}\label{eq:l2largrangian}
\mathbf{w}^*=\arg\min_{\mathbf{w}\in\mathcal{W}}\mathbf{w}'\boldsymbol{\Sigma}\mathbf{w} + \lambda \left\|\mathbf{w}\right\|_2^2,
\end{equation}
where $\lambda\geq 0$ is the penalty strength parameter and $\left\|\mathbf{w}\right\|_2^2 = \sum_{i=1}^N w_i^2$. Using the spectral decomposition of $\boldsymbol{\Sigma}=\mathbf{P}\boldsymbol{\Lambda}\mathbf{P}'$, where $\mathbf{P}\mathbf{P}'=\mathbb{I}_N$ and $\boldsymbol{\Lambda}=diag(\delta_1,\ldots,\delta_N)$, and since $\left\|\mathbf{w}\right\|_2^2=\mathbf{w}'\mathbf{w}=(\mathbf{P}'\mathbf{w})'(\mathbf{P}'\mathbf{w})$, we can rewrite the $\ell_2^2$ penalized objective function as
\begin{equation}\label{eq:l2meanvariance}
    \mathbf{w}^*=\arg\min_{\mathbf{w}\in\mathcal{W}} \mathbf{w}'\widetilde{\boldsymbol{\Sigma}}\mathbf{w},
\end{equation}
where $\widetilde{\boldsymbol{\Sigma}} = \mathbf{P}\left[\boldsymbol{\Lambda}+\lambda\mathbb{I}_N\right]\mathbf{P}'$ has all the eigenvalues shifted up by $\lambda\geq 0$. This is, again, a QP optimization problem that falls into our unified framework.

\subsection{Portfolio Optimization with $\ell_1$ Penalized Portfolio Norms}\label{sec:l1_constraint_portfolio_norms}

Similarly to the $\ell_2^2$-constraint, we can write the Lagrangian of $\ell_1$-constrained minimum-variance portfolio optimization problem as
\begin{equation}\label{eq:l1Lagrangian}
\mathbf{w}^*=\arg\min_{\mathbf{w}\in\mathcal{W}}\mathbf{w}'\boldsymbol{\Sigma}\mathbf{w} + \lambda \left\|\mathbf{w}\right\|_1,
\end{equation}
where $\lambda\geq 0$ is the penalty strength parameter and $\left\|\mathbf{w}\right\|_1 = \sum_{i=1}^N \left|w_i\right|$. The main difference compared to \eqref{eq:l2largrangian} is that the objective function in \eqref{eq:l1Lagrangian} is non-differentiable because of the kinks in the absolute value function, and the spectral decomposition will not help in converting \eqref{eq:l1Lagrangian} into a standard QP problem. Instead, we define $\mathbf{w}_+ = \max(0,\mathbf{w})\in\mathbb{R}_{0,+}^N$,
$\mathbf{w}_- = -\min(0,\mathbf{w})\in\mathbb{R}_{0,+}^N$
and $\mathbf{w}_+\cdot\mathbf{w}_-=\boldsymbol{0}$. Then $\mathbf{w}=\mathbf{w}_+-\mathbf{w}_-\mbox{ and } \left\|\mathbf{w}\right\|_1=(\mathbf{w}_+ + \mathbf{w}_-)'\mathbf{1}_N$.
We can rewrite the $\ell_1$-regularized objective function as
\begin{equation}\label{eq:l1meanvariance_old}
\left(\mathbf{w}^*,\mathbf{w}^*_+, \mathbf{w}^*_-\right)=\arg\min_{(\mathbf{w},\mathbf{w}_+,\mathbf{w}_-)\in\widetilde{\mathcal{W}}}\mathbf{w}'\boldsymbol{\Sigma}\mathbf{w} + \lambda \left(\mathbf{w}_+ + \mathbf{w}_-\right)'\boldsymbol{1}_N,
\end{equation}
where $\widetilde{\mathcal{W}}=\left\{(\mathbf{w},\mathbf{w}_+,\mathbf{w}_-)\in\mathbb{R}^{3N}: \mathbf{w}=\mathbf{w}_+-\mathbf{w}_-, \mathbf{w}_+\geq 0, \mathbf{w}_- \geq 0, \mbox{ and } \mathbf{w}\in\mathcal{W}\right\}.$ This way, we rewrote the original non-differentiable problem in $N$ variables as a QP problem in $3N$ variables with additional $N$ equality constraints.\footnote{It is possible to further simplify the optimization problem from $3N$ to $2N$ variables by incorporating these constraints explicitly. But we tested this empirically, and it slows down the algorithms because one needs to use then the $2N\times 2N$ matrix instead of $\boldsymbol{\Sigma}$ in \eqref{eq:l1meanvariance_old}. The same argument applies to the similar optimization problems below.} 

\begin{figure}
\centering
\subfloat[Long-only portfolio allocation\label{fig:allocation_long}]{%
\resizebox*{6.5cm}{!}{\includegraphics{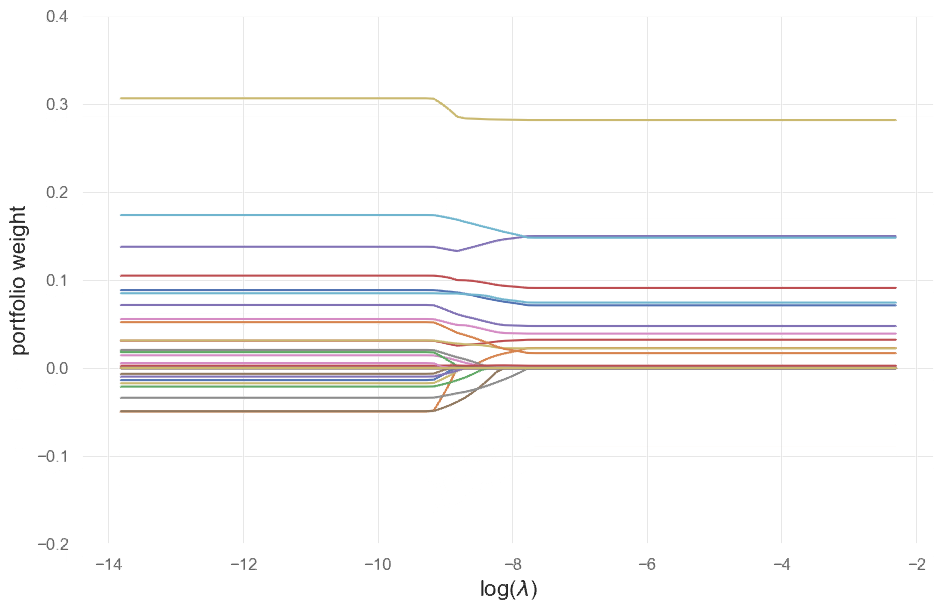}}}\hspace{5pt}
\subfloat[Long-short portfolio allocation\label{fig:allocation_longshort}]{%
\resizebox*{6.5cm}{!}{\includegraphics{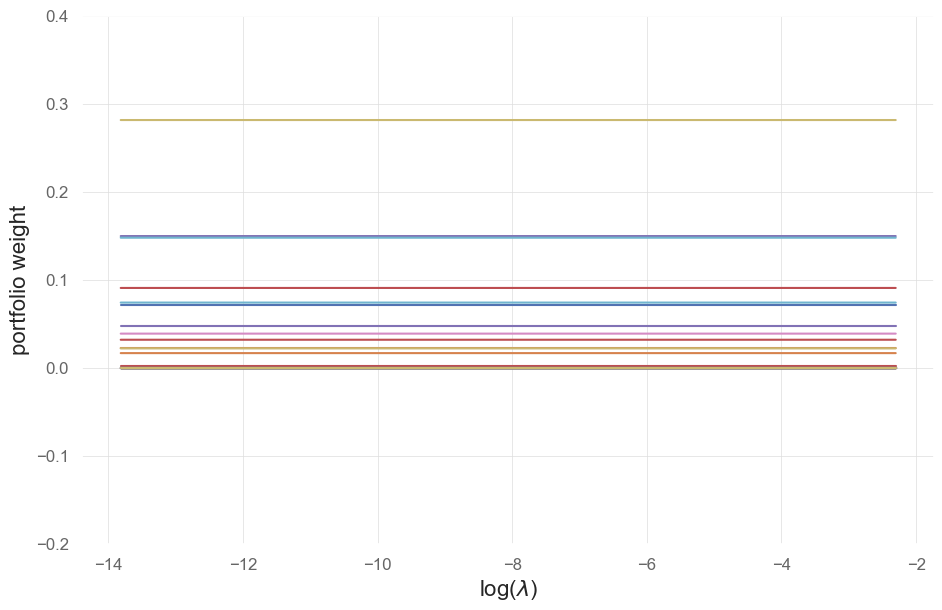}}}\hspace{5pt}
\caption{\protect \footnotesize Portfolio weights of N = 50 assets as a function of the regularization strength parameter $\lambda$ of $\ell_1$ penalty in minimum-variance $\ell_1$ regularized portfolio (long-only vs. long-short with $\vartheta=0.2$), the $x$-axis are in $\log$ scale.} \label{fig:allocation_l1}
\end{figure}

The following remarks can be made about this new optimization problem:
    \begin{itemize} 
        \item[(i)] Note that we do not have to include the constraint 
$\mathbf{w}_+\cdot\mathbf{w}_-=\boldsymbol{0}$ into the definition of the set of feasible weights $\widetilde{\mathcal{W}}$ since any portfolio with
$\mathbf{w}_+\cdot\mathbf{w}_-\neq \boldsymbol{0}$ is strictly dominated in terms of the value of the objective function by an analogous portfolio with 
$\mathbf{w}_+\cdot\mathbf{w}_-=\boldsymbol{0}$. Hence, the optimizer will never stop at $\mathbf{w}_+\cdot\mathbf{w}_- \neq \boldsymbol{0}$.
\item[(ii)] If the portfolio is long-only, the $\ell_1$ norm for the feasible portfolios reduces to the sum of portfolio weights, and the optimization problem \eqref{eq:l1Lagrangian} becomes differentiable. In this case, we observe empirically that optimal portfolio weights will never change when $\lambda$ grows---see the left panel in Figure \ref{fig:allocation_l1} (see also Figure \ref{fig:portfoliofrontier} where some optimal portfolios are $\ell_1$+$\ell_2^2$ regularized, and they coincide with the $\ell_2^2$ regularized portfolios). This is because the constraints will disappear if we assume that $\mathbf{w}^{\prime}\mathbf{1}_{N} = 1$ and $\mathbf{w} \geq \mathbf{0}_N$. Even when short positions are allowed, the optimization problem will have only \emph{partially} sparse solutions. In both cases, as opposed to a usual LASSO problem, the solution will not converge to $\mathbf{0}$ when $\lambda$ goes to infinity because we have another constraint in $\mathcal{W}$ that $\mathbf{w}'\mathbf{1}_N = 1$, and one will never get all the optimal weights equal to zero. As shown in the right panel in Figure \ref{fig:allocation_l1}, in the long-short portfolio, only all the initially (when $\lambda=0$) negative weights will converge to zero. Some of the initially positive weights will go to zero too. At the same time, the remaining positive weights will converge to a long-only minimum-variance portfolio. Importantly, some intermediate levels of $\lambda$ and the corresponding non-zero optimal weights can perform well out-of-sample.
\item[(iii)] Note that any of the constraints listed in Section \ref{sec:minimumvarianceportfolio} such that it involves an absolute value function, can be rewritten using the $\mathbf{w}^+$ and $\mathbf{w}^-$. Hence, the corresponding optimization problem can be solved using the QP methods.
\end{itemize}

\subsection{Portfolio Optimization with $\ell_1$+$\ell_2^2$ Penalized Portfolio Norms}\label{sec:l1_l2_regularization}

Naturally, we can consider both the $\ell_1$-constrained and $\ell_2^2$-constrained, which we call $\ell_1$+$\ell_2^2$-constrained portfolio. For that purpose, we  modify our objective function \eqref{eq:meanVariancePortOpt} to
\begin{equation}\label{eq:ela_Lagrangian}
\mathbf{w}^*=\arg\min_{\mathbf{w}\in\mathcal{W}}\mathbf{w}'\boldsymbol{\Sigma}\mathbf{w} + \lambda_1 \left\|\mathbf{w}\right\|_1 + \lambda_2 \left\|\mathbf{w}\right\|_2^2.
\end{equation}
By combining \eqref{eq:l2meanvariance} and \eqref{eq:l1meanvariance_old}, we can use again the eigenvalues decomposition of $\boldsymbol{\Sigma}=\mathbf{P}\boldsymbol{\Lambda}\mathbf{P}'$, where $\mathbf{P}\mathbf{P}'=\mathbb{I}_N$, $\boldsymbol{\Lambda}=diag(\delta_1,\ldots,\delta_N)$
and $\left\|\mathbf{w}\right\|_2^2=\mathbf{w}'\mathbf{w}=(\mathbf{P}'\mathbf{w})'(\mathbf{P}'\mathbf{w})$. 

\begin{equation}\label{eq:ela_meanvariance}
    \mathbf{w}^*=\arg\min_{(\mathbf{w},\mathbf{w}_+,\mathbf{w}_-)\in\widetilde{\mathcal{W}}} \mathbf{w}'\widetilde{\boldsymbol{\Sigma}}\mathbf{w}+\lambda \left(\mathbf{w}_+ + \mathbf{w}_-\right)'\boldsymbol{1}_N,
\end{equation}
where $\widetilde{\mathcal{W}}=\left\{(\mathbf{w},\mathbf{w}_+,\mathbf{w}_-)\in\mathbb{R}^{3N}: \mathbf{w}=\mathbf{w}_+-\mathbf{w}_-\mbox{ and } \mathbf{w}\in\mathcal{W}\right\}$ and $\widetilde{\boldsymbol{\Sigma}} = \mathbf{P}\left[\boldsymbol{\Lambda}+\lambda_2\mathbb{I}_N\right]\mathbf{P}'$ has shifted by $\lambda_2\geq 0$ all the eigenvalues.

\subsection{Long-Short Constrained Portfolio}\label{sec:long_short_portfolio}

The long-short constrained minimum-variance portfolio optimization from \eqref{eq:meanVariancePortOpt} is defined as
\begin{equation}\label{eq:longshortmeanvariance}
    \mathbf{w}^*(\vartheta)=\arg\min_{\mathbf{w}\in\mathcal{W}_{LS}(\vartheta)}\mathbf{w}'\boldsymbol{\Sigma}\mathbf{w},
\end{equation}
where $\mathcal{W}_{LS}(\vartheta)=\left\{\mathbf{w}\in\mathbb{R}^N:\sum_{i:w_i>0} w_i \leq 1+\vartheta\mbox{ and }\sum_{i:w_i<0} w_i\geq -\vartheta \right\}$. This is a different type of portfolio weights constraint that aggregates them based on their sign. Long-only portfolio constraint is a special case given by $\mathcal{W}_{LS}(\vartheta)$ for $\vartheta=0$. 
We can take again $\mathbf{w}_+ = \max(0,\mathbf{w})\in\mathbb{R}_{0,+}^N \mbox{ and }\mathbf{w}_- = -\min(0,\mathbf{w})\in\mathbb{R}_{0,+}^N$ and $\mathbf{w}_+\cdot\mathbf{w}_-=\boldsymbol{0}$. So, $\sum_{i:w_i>0} w_i \leq 1+\vartheta \iff \mathbf{w}_+'\boldsymbol{1}_N-1\leq \vartheta$ and  
$\sum_{i:w_i<0} w_i\geq -\vartheta \iff \mathbf{w}_-'\boldsymbol{1}_N\leq \vartheta$. 

Hence, we can replace the $\mathcal{W}_{LS}(\vartheta)$ with a new constraint set given by
$$\widetilde{\mathcal{W}}_{LS}(\vartheta)=\left\{\mathbf{w}\in\mathbb{R}^{N}:\mathbf{w} = \mathbf{w}_+ - \mathbf{w}_-, \mathbf{w}_+\geq 0, \mathbf{w}_- \geq 0, \mathbf{w}_+'\boldsymbol{1}_N\leq 1+ \vartheta, \mbox{ and }  \mathbf{w}_-'\boldsymbol{1}_N\leq \vartheta\right\}$$
and solve the corresponding QP problem.

\subsection{Mean-Variance Optimization with Risk-Free Asset}

In mean-variance portfolio in \eqref{eq:meanVariancePortOpt}, the goal is to optimize the trade-off between portfolio returns and risk. In other words, the mean-variance method looks for a portfolio with the lowest variance while the expected portfolio returns $\mathbf{w}^{\prime}\boldsymbol{\mu}$ is constraint from below by $\alpha_0$. Because of the convexity of the problem, the optimal value corresponds to the minimum volatility portfolio under the target return level.

In addition to the risky assets ($i=1,\ldots,N$) we can assume there is a risk-free asset for which $R_f = r_f, \mbox{ i.e., } \mathbb{E}[R_f] = r_f \mbox{ and } Var(R_f) = 0.$ Suppose the investor can invest in the $N$ risky investments as well as in the risk-free asset. The portfolio with investment in risk-free assets consists of two parts: $\mathbf{w}'\mathbf{1}_N = \sum_{i=1}^N w_i$ (invested in risky assets) and $1 - \mathbf{w}'\mathbf{1}_N$ (risk-free asset).

If borrowing is allowed, $(1-\mathbf{w}^{\prime}\mathbf{1}_N)$ can be negative. Long-short portfolio with return $R_{\mathbf{w}} = \mathbf{w}^{\prime}\mathbf{R} + (1-\mathbf{w}^{\prime}\mathbf{1}_N)R_f$ where $\mathbf{R}=[R_1,\ldots,R_N]^{\prime}$, has expected return 
$\mu_{\mathbf{w}} = \mathbf{w}^{\prime}\boldsymbol{\mu} + (1- \mathbf{w}^{\prime}\mathbf{1}_N)r_f$ and variance $\sigma_{\mathbf{w}} = \mathbf{w}^{\prime}\boldsymbol{\Sigma}\mathbf{w}$.

For a given choice of target mean return $\alpha_0$, choose the portfolio $\mathbf{w}^*$ to 
\begin{equation}\label{eq:meanvarianceLongShort}
    \mathbf{w}^*  = \arg\min_{\mathbf{w}^{\prime}\in\mathcal{W}} \frac{1}{2}\mathbf{w}^{\prime}\boldsymbol{\Sigma}\mathbf{w},
\end{equation}
 where $\mathcal{W}=\left\{\mathbf{w}\in \mathbb{R}^N: \mathbf{w}^{\prime}\boldsymbol{\mu} + (1-\mathbf{w}^{\prime}\mathbf{1}_N)r_f =\alpha_0\right\}$.
Then we can derive the Lagrangian as
\begin{equation}\label{eq:meanvariance_long-short_largrange}
    L(\mathbf{w},\lambda_1) = \frac{1}{2}\mathbf{w}^{\prime}\boldsymbol{\Sigma}\mathbf{w} - \gamma[(r_f - \alpha_0) + \mathbf{w}^{\prime}(\boldsymbol{\mu} - \mathbf{1}_N r_f)].
\end{equation}
Solving the Lagrangian, we get $\mathbf{w}^* = \gamma^* \boldsymbol{\Sigma}^{-1}(\boldsymbol{\mu} - \mathbf{1}_N r_f)$
and $\gamma^* = (\alpha_0 - r_f) / [(\boldsymbol{\mu} - \mathbf{1}_Nr_f)^{\prime}\boldsymbol{\Sigma}^{-1}(\boldsymbol{\mu} - \mathbf{1}_N r_f)]$. So the expected return and the variance of the optimal portfolio are given by $\mathbb{E}(R_N) = \mathbf{w}^{*\prime}\mathbf{R} + (1- \mathbf{w}^{*\prime}\mathbf{1}_N)r_f$,
$Var(R_N) = (\alpha_0 - r_f)^2/[(\boldsymbol{\mu} - \mathbf{1}_Nr_f)'\boldsymbol{\Sigma}^{-1}(\boldsymbol{\mu}-\mathbf{1}_Nr_f)]$, respectively.

Note that because of the risk-free asset, the resulting portfolio frontier will be a line (it is the so-called one fund theorem) connecting two points in the mean-variance plane: the $(0,r_f)$ where all the money is invested only in the risk-free asset; and the mean and variance of so called market portfolio $\mathbf{w}_0 = \boldsymbol{\Sigma}^{-1}(\boldsymbol{\mu} - \mathbf{1}_N r_f)/[\mathbf{1}'\boldsymbol{\Sigma}^{-1}(\boldsymbol{\mu} - \mathbf{1}_N r_f)]$ which is the tangent point to the portfolio frontier without the risk-free asset. So in order to find solutions for different $\alpha_0$,  it suffices to solve for the portfolio without risk-free asset, and take linear combinations of that portfolio with the risk-free investment. Hence, again this can be considered as part of our general portfolio framework.

\subsection{Maximum Sharpe Ratio Portfolio}\label{sec:maxSharpeRatioPortfolio}

Markowitz’s mean-variance framework in \eqref{eq:meanVariancePortOpt} provides portfolios along the optimal frontier, and the choice of the specific portfolio depends on the risk-aversion of the investor. Typically one measures the investment performance using the Sharpe ratio, and there is only one portfolio on the optimal frontier that achieves the maximum Sharpe ratio
\begin{equation}\label{eq:maximumSharperatio}
\arg\max_{\mathbf{w}\in\mathcal{W}}\frac{\mathbf{w}'\boldsymbol{\mu} - r_f}{\sqrt{\mathbf{w}'\boldsymbol{\Sigma}\mathbf{w}}},
\end{equation}
where $\mathcal{W}=\left\{\mathbf{w}\in\mathbb{R}^N: \mathbf{1}_N\mathbf{w}=1, \mathbf{w}\geq \boldsymbol{0}\right\}$, and $r_f$ is the return for a risk-free asset.

This problem -- although nonconvex -- belongs to the family of so called Fractional Programming (FP) optimization problems that involve ratios. It is a concave-convex single-ratio and can be solved by different approaches. This particular FP problem is still simple to solve using a reparametrization trick. One can note that the objective function in \eqref{eq:maximumSharperatio} is homogeneous of degree zero, and reformulate this problem as a QP problem. If there exists at least one portfolio vector  $\mathbf{w}$ such that $\mathbf{w}'\boldsymbol{\mu} - r_f> 0$, then for $\mathbf{w}'\boldsymbol{\mu} - r_f\neq 0$, and $\mathbf{w}\in\mathcal{W}$, we can change the maximization problem into an equivalent minimization
\begin{equation}\label{eq:maximumSharperatio_transformed}
\arg\min_{\mathbf{w}\in\mathcal{W}}\frac{\sqrt{\mathbf{w}'\boldsymbol{\Sigma}\mathbf{w}}}{\mathbf{w}'(\boldsymbol{\mu} - r_f\mathbf{1}_N)},
\end{equation}
where $\mathcal{W}=\left\{\mathbf{w}\in\mathbb{R}^N: \mathbf{w}'\mathbf{1}_N=1, \mathbf{w}\geq \boldsymbol{0}\right\}$. Now by the homogeneity of degree zero of the objective function, we can choose the proper scaling factor for our convenience. We define $\widetilde{\mathbf{w}}=\gamma \mathbf{w}$ with scaling factor $\gamma = 1/\mathbf{w}'(\boldsymbol{\mu}-r_f\mathbf{1}_N)>0$. So that the objective becomes $\widetilde{\mathbf{w}}'\boldsymbol{\Sigma}\widetilde{\mathbf{w}}$, the sum constraint $\mathbf{1}_N'\widetilde{\mathbf{w}}=\gamma$, and the above problem is equivalent to 
\begin{equation}\label{ep:maximumShaperation_final}
\arg\min_{\mathbf{w}\in\mathcal{W}}\frac{\sqrt{\gamma\mathbf{w}'\boldsymbol{\Sigma}\mathbf{w}\gamma}}{\gamma\mathbf{w}'(\boldsymbol{\mu} - r_f\mathbf{1}_N)} \iff \arg\min_{[\widetilde{\mathbf{w}},\gamma]'\in\widetilde{\mathcal{W}}}\widetilde{\mathbf{w}}'\boldsymbol{\Sigma}\widetilde{\mathbf{w}},
\end{equation}
where $\widetilde{\mathcal{W}}=\left\{[\widetilde{\mathbf{w}},\gamma]'\in\mathbb{R}^{N+1}: 1 = \widetilde{\mathbf{w}}'(\boldsymbol{\mu} - r_f\mathbf{1}_N), \mathbf{1}_N'\widetilde{\mathbf{w}}=\gamma, \widetilde{\mathbf{w}}\geq \boldsymbol{0}\right\}$.
 
The optimal portfolio weights $\mathbf{w}^*$  are recovered after doing the optimization through the transformation $\mathbf{w}^*=\widetilde{\mathbf{w}}^*/\gamma^*$.
Importantly note that all the aforementioned constraints and regularizations can also be incorporated into this optimization problem  \eqref{ep:maximumShaperation_final}, and it will remain equivalent to the original maximum Sharpe ratio portfolio with the same regularizations and constraints properly rescaled as in \eqref{eq:maximumSharperatio_l1l2}. In Section \ref{sec:maxSharpeRatioPortfolio_PCAbased}, we will provide a more detailed and precise presentation. The advantage of \eqref{ep:maximumShaperation_final} is that even with these constraints and regularizations, it will be easy to solve numerically using QP methods.

\begin{table}[ht]
\centering
\caption{\protect \footnotesize \textit{Summary of the total running time (in seconds) for $100$ rolling windows of three different portfolio optimization problems from our general framework described in Section \ref{sec:portfolio} for different dimensions of the problem ($N=10,20,50,500$), and two different levels of tolerance and precision in the optimizer: (i) default precision used in the OSQP package \protect\url{https://osqp.org} ; (ii) high precision with $10^{4}$ maximum iterations, and the absolute and relative tolerance set to $10^{-8}$. The latter is needed to generate convex portfolio frontiers in simulations for large $N$, and we use it in all our empirical studies. Computations are done using a single core of the AMD Ryzen Threadripper 2990WX Processor.}}
\label{table:time_table}
\resizebox{\columnwidth}{!}{%
\begin{tabular}{lrrrr|rrrr}
\hline
 &     \multicolumn{4}{c}{{Default Precision}} & \multicolumn{4}{c}{{High Precision}} \\
 & & & & & & & &\\
{Portfolio  Objective Function}&  N=10 &      N=20 &      N=50 &     N=500 & N=10 &      N=20 &      N=50 &     N=500 \\
\hline
Long-Only Min-Variance         &  0.14 &  0.16 &  0.20 &  10.77  &  0.15 &  0.16 &  0.21 &  11.96\\
& & & & \\
Long-Only Max-Sharpe Ratio          &  0.12 &  0.13 &  0.21 &  29.94  &  0.17 &  0.19 &  0.26 &  63.02 \\
& & & & \\
Long-Short Max-Sharpe Ratio with $\ell_1$$+\ell_2^2$  &  0.17 &  0.21 &  0.45 &  62.64 &  0.21 &  0.26 &  0.53 &  222.17\\
\hline
\end{tabular}
}
\end{table}

In our portfolio optimization framework, once the portfolio problems are turned into standard QP problems, we use the OSQP solver from \citet{osqp} to solve them. The solver uses ADMM algorithm for the optimization (see \citealp{boydADMM:2011} and references therein for the detail introduction of the algorithm). It is an open-source solver available at \url{https://osqp.org/docs/solver/index.html}. As summarized in Table \ref{table:time_table}, portfolios with 50 assets or less can be optimized with very high precision especially compared to any numerical gradient based method. All the evaluations in Table \ref{table:time_table} are done on a single core of the AMD Ryzen Threadripper 2990WX Processor. This concludes our summary of portfolio optimization problems that we can solve using the QP framework. The corresponding code with the implementation in Python is available online at 
\url{https://github.com/PawPol/PyPortOpt}. We describe next all the covariance matrix estimators considered in this paper.

\section{Modeling Stock Returns}\label{sec:covariance_model}

In Markowitz's portfolio theory, the mean vector $\boldsymbol{\mu}$ and the covariance matrix $\boldsymbol{\Sigma}$ are assumed to be known. However, in practice, these parameters must be estimated from data. A prevalent method involves using the historical sample mean and sample covariance matrix under the assumption of $iid$ observations. This approach frequently results in suboptimal out-of-sample performance. As highlighted in the introduction, there exist alternative estimators that offer improved out-of-sample outcomes. In the subsequent empirical section, we utilize our portfolio optimization framework to compare the portfolio performance yielded by various mean and covariance matrix shrinkage methodologies against that from different factor-based models. The former, the shrinkage methods, derive their estimates from daily data, while the latter, the factor-based models, utilize monthly returns and stock specific characteristics for their evaluations.

In case of daily data and the mean and covariance matrix shrinkage, for the mean estimation we use the sample mean and three shrinkage estimators from \citet{wang2014non} and  \citet{bodnar2019optimal}. For the covariance matrix, first, we use the classical linear shrinkage covariance matrix estimator \citet{Ledoit:04} defined as
\begin{equation}
    \hat{\boldsymbol{\Sigma}} = \hat{\delta}\hat{\mathbf{F}} + (1-\hat{\delta})\mathbf{S},
\end{equation}
where $\mathbf{S} =\frac{1}{T} \sum_{t=1}^T(\mathbf{r}_t - \Bar{\mathbf{r}})(\mathbf{r}_t - \Bar{\mathbf{r}})^{\prime}$ and $\hat{\mathbf{F}}$ is the estimated structured covariance matrix. In particular, $\hat{\mathbf{F}} = trace(S)/N$, and $\hat{\delta}$ denotes the estimator of optimal shrinkage constant $\delta$. In practice, the authors propose to use $\hat{\delta} = \max\{0, \min\{\frac{\hat{\kappa}}{T}, 1\}\}$, where $\hat{\kappa} = \frac{\hat{\pi}-\hat{\rho}}{\hat{\gamma}}$, and $\hat{\pi}, \hat{\rho}$ and $\hat{\gamma}$ be estimated as $\hat{\pi} = \sum_{i=1}^N\sum_{j=1}^N\hat{\pi}_{ij}$ with $ \hat{\pi}_{ij} = \frac{1}{T}\sum_{t=1}^T\{(r_{it}-\Bar{r}_{i.})(r_{jt}-\Bar{r}_{j.}) - s_{ij}\}$, $\hat{\rho} = \sum_{i=1}^N\hat{\pi}_{ij} + \sum_{i=1}^N\sum_{j=1,j\neq i}^N\frac{\Bar{r}}{2}(\sqrt{\frac{s_{jj}}{s_{ii}}}\hat{\theta}_{ii,ij} + \sqrt{\frac{s_{ii}}{s_{jj}}}\hat{\theta}_{jj,ij})$ with $\hat{\theta}_{ii,ij} = \frac{1}{T}\sum_{t=1}^T\{(r_{it}-\Bar{r}_{i.})^2-s_{ii}\}\{(r_{it}-\Bar{r}_{i.})(r_{jt}-\Bar{r}_{j.}) - s_{ij}\}$, $\hat{\theta}_{jj,ij} = \frac{1}{T}\sum_{t=1}^T\{(r_{jt}-\Bar{r}_{j.})^2-s_{jj}\}\{(r_{it}-\Bar{r}_{i.})(r_{jt}-\Bar{r}_{j.}) - s_{ij}\}$, and $\hat{\gamma} = \sum_{i=1}^N\sum_{j=1}^N(f_{ij} - s_{ij})^2$.

In situations when the number of assets (variables) is commensurate with the sample size, the sample covariance matrix is usually not well-conditioned and not invertible. Getting the linear combination of the sample covariance matrix and identity matrix is a way to shrink the eigenvalues of the sample covariance matrix away from zero and towards their average in  $\hat{\mathbf{F}} = trace(S)/N$, with $\delta\in[0,1]$ denoting the shrinkage intensity. As a result, we get a well-conditioned covariance matrix estimator that has a lower mean-square error than the sample covariance matrix, and, in large dimensions, when $N$ grows asymptotically with $T$, it is a consistent estimator of the covariance matrix.

Second, we consider a more recent nonlinear shrinkage covariance matrix estimator---the quadratic inverse shrinkage estimator from \citet{ledoit2020analytical}. The estimator can be written as
\begin{equation}
    \hat{\boldsymbol{\Sigma}}_t := \mathbf{U}_t\hat{\boldsymbol{\Delta}}_t\mathbf{U}_t^{\prime},
\end{equation}
where $\hat{\boldsymbol{\Delta}}_t := diag(\hat{\delta}_t(\lambda_{1,t}), \ldots, \hat{\delta}_t(\lambda_{N,t}))$, and $\hat{\delta}_t$ is a real univariate function of $\lambda_{i,t}$ for $i = 1,\ldots, N$. $\boldsymbol{\lambda} = (\lambda_1, \ldots, \lambda_N)$ denotes the eigenvalues and $\mathbf{U}_t = [u_{1,t}, \ldots, u_{N,t}]$ are the corresponding eigenvectors. By introducing the nonlinear transformation (Hilbert transform) of the sample eigenvalues, this method helps with the curse of dimensionality. 

The shrinkage techniques previously described are typically employed for large-dimensional portfolio problems. A different strategy to address the challenges of dimensionality in portfolio optimization involves the use of factor models. Classical factor modeling, as presented by \cite{FAMA1993}, \cite{carhart1997persistence}, and \cite{fama2015five}, assumes that returns adhere to the linear model:
\begin{equation} \label{eq:Fama_French_Factor_Model}
r_{i, t} = \alpha_{i} + \boldsymbol{\beta}{i}' \mathbf{f}{t} + \epsilon_{i, t},
\end{equation}
where $\mathbf{f}t \in \mathbb{R}^{K \times 1}$ represents a vector of observed factors, $\epsilon{i,t}$ is the zero-mean noise that captures the idiosyncratic component uncorrelated with the observed factors, and $\boldsymbol{\beta}{i} \in \mathbb{R}^{K\times 1}$ denotes a vector of unknown factor loadings. In many of these models, $\alpha_{i}$ is set to $0$ for all assets $i$. Given that this is essentially a linear regression problem and the factors are presumed to be uncorrelated with $\epsilon_{i,t}$, the return's covariance matrix divides into a section explained by the factors and an idiosyncratic section. Additionally, if the $\epsilon_{i,t}$ components are assumed to be uncorrelated across assets, the covariance matrix of the idiosyncratic component can be directly estimated from the regression residuals. Consequently, this model remains applicable even when $N$ significantly exceeds $T$.

However, the factor model given in \eqref{eq:Fama_French_Factor_Model} has its limitations. First, it assumes that the factors are both known and common across all assets. This means they can only elucidate risk to a certain extent and may not always correlate strongly with the actual risk in specific market conditions. Second, the factor loadings, represented by $\boldsymbol{\beta}_{i}$, are considered constant over time.

An alternative method that addresses the first limitation is to employ Principal Component Analysis (PCA) to derive latent factors directly from the covariance matrix of asset returns, without needing additional information. However, the covariance matrix for individual stock returns does not possess a lower-dimensional latent subspace that can precisely capture the variations in these returns. As a consequence, executing PCA on the covariance matrix of individual stock returns tends to introduce significant noise. This can lead to unstable portfolios and underperformance in out-of-sample scenarios. Thus, rather than applying PCA directly to the matrix of stock returns, it's more effective to work with the matrix of returns from portfolios that are single or double-sorted based on a cross-section of firm characteristics, as discussed in \citet{bryzgalova2020forest} and the references therein. 

PCA, when applied to managed portfolios, can extract factors that encapsulate the co-movement among returns and identify systematic time-series factors that predominantly influence cross-sectional risk. Typically, the top \( K \) eigenvectors are selected as assets in the portfolios, and one then optimizes the best capital allocation among them. \citet{lettau2020factors} introduce the Risk Premium (RP)-PCA that identifies pivotal factors in explaining asset returns. While traditional PCA focuses solely on data comovement, it does not incorporate data means. Consequently, it may miss out on capturing vital differences in the mean risk premia of assets.  In contrast, RP-PCA takes into account both the first and second moments of data, thereby enhancing estimation efficiency. Our empirical results confirm that RP-PCA outperforms PCA in portfolio performance.

\citet{bryzgalova2020forest} introduced the so-called Asset Pricing (AP) Trees, which serve as a generalization of sorting portfolios using tree-based methods. AP Trees offer concise and interpretable portfolios that span the stochastic discount factor (SDF) on stock returns; and it addresses challenges related to complexity, high dimensionality, and duplication. Our empirical analysis employs excess returns from AP-Trees with depth equal to three, as well as a broad cross-section of single-sorted decile portfolios. These portfolios are derived from ten distinct deciles of 33 anomaly characteristics, resulting in a total of 330 managed portfolios for the single sorting. We do not work with double sorted portfolios because in our universe of mid- and large-cap stocks considered in the empirical analysis many of the double sorted portfolios were empty. AP-Trees approach results in 36 different sortings---out of 10 stock specific characteristics we always use Size, and remaining two (9 choose 2) give 36 different trees of depth three. Each of these trees comprises 360 managed portfolios.

The AP-Trees and all the PCA-based models still assume static loadings, and they lack accuracy and flexibility because after constructing the managed portfolios, they use only the information from their returns to estimate optimal portfolio positions. In a similar way \citet{KELLY2019501} motivated their IPCA model, where asset returns are assumed to admit the following factor structure

\begin{equation}\label{eq:IPCA_model}
     r_{i, t+1} = \alpha_{i,t} + \boldsymbol{\beta}'_{i,t} \mathbf{f}_{t+1} + \epsilon_{i, t+1}, \quad \forall i=1,\ldots,N \mbox{ and } t=1,\ldots,T.
\end{equation}

The major distinctions from the classical factor models discussed previously are:
\begin{itemize}
\item[(i)] The IPCA model, analogous to BARRA's factor model, posits that the alphas $\alpha_{i,t}$ and the factor loadings $\boldsymbol{\beta}_{i,t} \in \mathbb{R}^{K\times 1}$ are time-dependent. However, unlike BARRA's model, it assumes they are implicitly observed through
\[\alpha_{i,t} = \mathbf{z}_{i,t}^{\prime}\boldsymbol{\Gamma}_{\alpha} + v_{\alpha,i,t}, \quad \boldsymbol{\beta}_{i,t} = \mathbf{z}_{i,t}^{\prime}\boldsymbol{\Gamma}_{\beta} + \mathbf{v}_{\beta,i,t},\]
where $\mathbf{z}_{i,t}\in \mathbb{R}^{1\times L}$ denotes observed asset-specific characteristics, and 
$\boldsymbol{\Gamma}_{\alpha}\in \mathbb{R}^{L\times 1}$ and $\boldsymbol{\Gamma}_{\beta}\in \mathbb{R}^{L\times K}$ are matrices of parameters estimated from the data.
\item[(ii)] Due to the dimension reduction introduced by the matrix $\boldsymbol{\Gamma}_{\beta}\in \mathbb{R}^{L\times K}$, the number of observed factors $L$ can be much larger than the number of factor loadings $K$.
\item[(iii)] The factors $\mathbf{f}_t\in \mathbb{R}^{K\times 1}$ are time-dependent and are estimated from the data.
\item[(iv)] This model is predictive, with observable factors lagged by one period relative to the returns they explain.
\item[(v)] $\epsilon_{i, t+1}$, $v_{\alpha,i,t}$, and $\mathbf{v}_{\beta,i,t}$ are mean zero random noises originating from the estimation of factors and loadings. The $\epsilon_{i, t+1}$ uncovers the firm-level risk, whereas $v_{\alpha,i,t}$ and $\mathbf{v}_{\beta,i,t}$ represent the residuals between the true factor model parameters and observable firm characteristics.
\end{itemize}

The rationale behind the IPCA model lies in the challenge of high-dimensional factor models: an excess of characteristics can lead to significant noise and collinearity among factors. This makes the results challenging to interpret and can diminish the model's out-of-sample performance. Hence, $\boldsymbol{\Gamma}_{\beta}$ is introduced to aggregate large-dimensional characteristics into a linear combination of exposure risks. Any errors orthogonal to the dynamic loadings are accounted for in the $\mathbf{v}_{\beta,i,t}$.

In the empirical analysis, we assume that $\boldsymbol{\Gamma}_{\alpha} = \mathbf{0}$ while focusing on the estimation of $\boldsymbol{\Gamma}_{\beta}$. Hence, for the restricted model ($\boldsymbol{\Gamma}_{\alpha} = \mathbf{0}$), we have
\begin{equation}
    r_{i, t+1} =  \mathbf{z}_{i,t}^{\prime}\boldsymbol{\Gamma}_{\beta} \mathbf{f}_{t+1} + \epsilon_{i, t+1}^{*},
\end{equation}
where $\epsilon_{i, t+1}^{*} = \epsilon_{i, t+1} +v_{\alpha,i,t}+\mathbf{v}_{\beta,i,t}\mathbf{f}_{t+1}$. We can derive this based on the vector form
\[ \mathbf{r}_{t+1} =  \mathbf{Z}_{t}^{\prime}\boldsymbol{\Gamma}_{\beta} \mathbf{f}_{t+1} + \boldsymbol{\epsilon_{t+1}^{*}}, \]
where $\boldsymbol{r_{t+1}}$ is an $N\times 1$ vector of assets returns, $\mathbf{Z}_t$ is an $N\times L$ vector of observable characteristics and $\boldsymbol{\Gamma}_{\beta}$ is an $L\times K$ mapping matrix, $\mathbf{f}_{t+1}$ is an $K\times 1$ vector of the combination latent factor. Then we can write the objective function of IPCA model as
\begin{equation}\label{eq: IPCA_objective}
    \min_{\boldsymbol{\Gamma}_{\beta}, F}\sum_{t=1}^{T-1}(\mathbf{r}_{t+1} - \mathbf{Z}_{t}^{\prime}\boldsymbol{\Gamma}_{\beta} \mathbf{f}_{t+1})^{\prime}(\mathbf{r}_{t+1} - \mathbf{Z}_{t}^{\prime}\boldsymbol{\Gamma}_{\beta} \mathbf{f}_{t+1}),
\end{equation}
with constrain $\boldsymbol{\Gamma}_{\beta}^{\prime}\boldsymbol{\Gamma}_{\beta} = \boldsymbol{I}_k$ and $\mathbf{FF}^{\prime} = diag(\lambda_1, \ldots, \lambda_k)$. To minimize the objective function \eqref{eq: IPCA_objective}, one iterates
\begin{subequations}\label{eq:IPCA_iterate}
    \begin{equation}\label{eq:IPCA_latent_factor}
        \hat{\mathbf{f}}_{t+1} = (\hat{\boldsymbol{\Gamma}}_{\beta}^{\prime}\mathbf{Z}_t^{\prime}\mathbf{Z}_t\hat{\boldsymbol{\Gamma}}_{\beta})^{-1}\hat{\boldsymbol{\Gamma}}_{\beta}^{\prime}\mathbf{Z}_t^{\prime}\mathbf{r}_{t+1}, \quad \text{for all} \,\, t,
    \end{equation}
 \mbox{and}   \begin{equation}\label{eq:IPCA_dimensional_reduction}
        \mbox{vec}(\hat{\boldsymbol{\Gamma}}_{\beta}^{\prime}) = (\sum_{t=1}^{T-1}\mathbf{Z}_t^{\prime}\mathbf{Z}_t\otimes \hat{\mathbf{f}}_{t+1}\hat{\mathbf{f}}_{t+1}^{\prime})^{-1}(\sum_{t=1}^{T-1}[\mathbf{Z}_t\otimes\hat{\mathbf{f}}_{t+1}^{\prime}]^{\prime}\mathbf{r}_{t+1}),
    \end{equation}
\end{subequations}
where $\otimes$ denotes the Kronecker product of matrices. Formula \eqref{eq:IPCA_latent_factor} shows that latent factors represent the coefficients of returns regressed on the latent loading matrix $\boldsymbol{\beta}_t \in \mathbb{R}^{N \times L}, t = (1, \ldots, T)$. Meanwhile, $\boldsymbol{\Gamma}_{\beta}$ denotes the regression coefficients of $\mathbf{r}_{t+1}$ on the combination of latent factors and firm characteristics. This first-order condition system does not have a close form solution, but it can be solved numerically by the alternating least squares method. 

\section{Regularizing Factor-Based Portfolios: An Application to the Maximum Sharpe Ratio Objective}\label{sec:maxSharpeRatioPortfolio_PCAbased}

In portfolio optimization, among all the objective functions in our framework, we focus on two fully-invested optimal portfolios: the minimum variance (min Var) portfolio, as detailed in Section \ref{sec:minimumvarianceportfolio}, and the maximum Sharpe ratio (max SR) portfolio, discussed in Section \ref{sec:maxSharpeRatioPortfolio}. We consider both with and without the $\ell_1+\ell_2^2$ regularization, which is covered in Section \ref{sec:l1_l2_regularization}. The minimum variance portfolio is commonly employed to evaluate models that emphasize covariance matrix estimation without mean prediction. In our study, we use it for daily data, specifically for all covariance matrix shrinkage models and for the $\ell_1+\ell_2^2$ regularized portfolio problems. On the other hand, the maximum Sharpe ratio portfolio aims to maximize the risk-adjusted return of the portfolio strategy, meaning it offers the highest return for each unit of risk, measured in terms of portfolio volatility. Positioned centrally on the portfolio efficient frontier, it is one of the most computationally intensive problems in our framework, as it necessitates reparametrization into a higher dimensional space. Therefore, we consider it a good representative for our mean (and covariance matrix) shrinkage models using daily returns, as well as for the factor-based models employing monthly returns, given the persistence of the mean signal in the constructed factor portfolios. The corresponding optimization problem can be expressed as:

\begin{align}\label{eq:maximumSharperatio_used}
&\arg\max_{\mathbf{w}\in\mathcal{W}_{LS}(\vartheta, \mathbf{V})}\frac{\mathbf{w}'\widehat{\boldsymbol{\mu}} - r_f}{\sqrt{\mathbf{w}'\widehat{\boldsymbol{\Sigma}}\mathbf{w}}},\\
\mathcal{W}_{LS}(\vartheta, \mathbf{V})=&\left\{\mathbf{w}\in\mathbb{R}^K:  (\mathbf{V}\mathbf{w})'\mathbf{1}_N=1, L_j\leq (\mathbf{V}\mathbf{w})_j\leq U_j, \forall j=1,\ldots,N  \right.\nonumber \\
& \left. \sum_{i:(\mathbf{V}\mathbf{w})_i>0} (\mathbf{V}\mathbf{w})_i \leq 1+\vartheta, \sum_{i:(\mathbf{V}\mathbf{w})_i<0} (\mathbf{V}\mathbf{w})_i\geq -\vartheta\right\}\nonumber 
\end{align}

where $\vartheta\geq 0$ represents the short-selling threshold parameter (set to $\vartheta=0.2$ in our study). The matrix $\mathbf{V}\in\mathbb{R}^{N\times K}$ encapsulates the linear mapping between managed portfolios and individual assets in our investment universe. We set the other parameters as follows: $r_f = 0$, $L_j=-0.08$, and $U_j=0.08$ for all $j=1,\ldots,N$. If the optimal portfolio weights $\mathbf{w}$ pertain to individual stocks, then $\mathbf{V}$ is the identity matrix, and $\widehat{\boldsymbol{\mu}}$ and $\widehat{\boldsymbol{\Sigma}}$ signify the mean and covariance matrix (after shrinkage) estimators of those individual stock returns. For AP-Trees, we employ the high-dimensional sample mean and covariance matrix of factor portfolios. With PCA-based models, we consider $K=2,\ldots,6$ dimensional $\widehat{\boldsymbol{\mu}}_f$, and $\widehat{\boldsymbol{\Sigma}}_f$ derived from PCA, RP-PCA, and IPCA estimated means, along with the estimated covariance matrix of the corresponding $K$ factor portfolios. For PCA and RP-PCA, $\mathbf{V}$ comprises the first $K$ eigenvectors of the PCA and RP-PCA covariance matrices, respectively. In the case of the IPCA model, $\mathbf{V} = (\hat{\boldsymbol{\Gamma}}_{\beta}^{\prime}\mathbf{Z}_t^{\prime}\mathbf{Z}_t\hat{\boldsymbol{\Gamma}}_{\beta})^{-1}\hat{\boldsymbol{\Gamma}}_{\beta}^{\prime}\mathbf{Z}_t^{\prime}$ describes the transformation from the IPCA factors of the last observation to individual stocks.

In order to solve it efficiently, we reformulate \eqref{eq:maximumSharperatio_used} into an equivalent QP problem from Section \ref{sec:maxSharpeRatioPortfolio} with constraints rewritten as in Section \ref{sec:long_short_portfolio}. In Section \ref{sec:empirics}, we introduce factor portfolios based on Principal Component Analysis (PCA), Risk Premium PCA (RP-PCA), and Instrumented PCA (IPCA). All these PCA-based models correspond to low-dimensional portfolio problems with $\mathbf{w}\in \mathbb{R}^K$. If we were to continue applying the $\ell_1$ and $\ell_2$ penalties to each factor, it would not yield a sparse solution for either the managed portfolios or the individual stock weights. Therefore, for the PCA-based models, we define an $\ell_1+\ell_2^2$ regularized maximum Sharpe ratio portfolio as:
\begin{equation}\label{eq:maximumSharperatio_l1l2}
\color{black}
\arg\min_{\mathbf{w}\in\mathcal{W}_{LS}(\vartheta,\mathbf{V})}\frac{-(\mathbf{w}'\boldsymbol{\mu}_f - r_f)}{\sqrt{\mathbf{w}'\boldsymbol{\Sigma}_f\mathbf{w}}} + {\delta}_1 \frac{1}{\mathbf{w}^T\boldsymbol{\mu}_f - r_f} \left\|\mathbf{V}\mathbf{w}\right\|_1 + {\delta}_2 \frac{1}{\left(\mathbf{w}'\boldsymbol{\mu}_f - r_f\right)^2} \left\|\mathbf{V}\mathbf{w}\right\|_2^2,
\end{equation}
where $\mathcal{W}_{LS}(\vartheta, \mathbf{V})$ is the same as in \eqref{eq:maximumSharperatio_used}. Depending on the choice of $\mathbf{V}$, the regularization terms in \eqref{eq:maximumSharperatio_l1l2} are with respect to the managed portfolios (in PCA and RP-PCA) or the individual stocks (in IPCA). Next, we reparametrize the optimization problem in \eqref{eq:maximumSharperatio_l1l2} as
\begin{equation}\label{eq:MaximumSharperatio_reparameterize}
\color{black}
\arg\min_{[\mathbf{w},\gamma]'\in\widetilde{\mathcal{W}}_{LS}(\vartheta, \mathbf{V})}-\frac{\gamma\mathbf{w}'(\boldsymbol{\mu}_f - r_f\mathbf{1}_K)}{\sqrt{\gamma\mathbf{w}'\boldsymbol{\Sigma}_f\mathbf{w}\gamma}} + \delta_1\left\|\gamma\mathbf{Vw}\right\|_1 + \delta_2\left\|\gamma\mathbf{Vw}\right\|_2^2,
\end{equation}
subject to an additional constraint $\gamma = 1/\mathbf{w}'(\boldsymbol{\mu}_f-r_f\mathbf{1}_K)$. Now, by defining $\widetilde{\mathbf{w}} = \gamma \mathbf{w}$, we obtain the corresponding quadratic programming problem
\begin{equation}\label{eq:MaximumSharperatio_model}
\color{black}
\arg\min_{[\widetilde{\mathbf{w}},\gamma]'\in\widetilde{\mathcal{W}}_{LS}(\vartheta, \mathbf{V})}{\widetilde{\mathbf{w}}'\boldsymbol{\Sigma}_f\widetilde{\mathbf{w}}} + \lambda_1\left\|\mathbf{V}\widetilde{\mathbf{w}}\right\|_1 + \lambda_2\left\|\mathbf{V}\widetilde{\mathbf{w}}\right\|_2^2,
\end{equation}
\begin{align*}
\text{where }&\widetilde{\mathcal{W}}_{LS}(\vartheta, \mathbf{V}) = \{ [\widetilde{\mathbf{w}},\gamma]' \in \mathbb{R}^{(K+1)} : (\mathbf{V}\widetilde{\mathbf{w}})'\mathbf{1}_N = \gamma, \\
& \sum_{i: (\mathbf{V}\widetilde{\mathbf{w}})_i > 0} (\mathbf{V}\widetilde{\mathbf{w}})_i \leq \gamma(1 + \vartheta), \sum_{i: (\mathbf{V}\widetilde{\mathbf{w}})_i < 0} (\mathbf{V}\widetilde{\mathbf{w}})_i \geq -\gamma \vartheta,\ L_j\gamma \leq (\mathbf{V}\widetilde{\mathbf{w}})_j \leq U_j\gamma,\  \forall j \}.
\end{align*} Similarly to \eqref{eq:l2meanvariance} and \eqref{eq:l1meanvariance_old}, we can employ the eigenvalues decomposition of $\boldsymbol{\Sigma_f}=\mathbf{P}\boldsymbol{\Lambda_f}\mathbf{P}'$, where $\mathbf{P}\mathbf{P}'=\mathbb{I}_K$, $\boldsymbol{\Lambda_f}=\text{diag}(\delta_1,\ldots,\delta_K)$, and $\left\|\mathbf{V}\widetilde{\mathbf{w}}\right\|_2^2=(\mathbf{V}\widetilde{\mathbf{w}})'\mathbf{V}\widetilde{\mathbf{w}}=(\mathbf{P}'\mathbf{V}\widetilde{\mathbf{w}})'(\mathbf{P}'\mathbf{V}\widetilde{\mathbf{w}})$  to reduce \eqref{eq:MaximumSharperatio_model} to
\begin{equation}\label{eq:ela_factor_modified}
    \color{black}
\arg\min_{(\widetilde{\mathbf{w}},\gamma,\mathbf{v}_+,\mathbf{v}_-)\in\widetilde{\mathcal{W}}_{LS}^{\pm}(\vartheta,\mathbf{V})} \widetilde{\mathbf{w}}'\widetilde{\boldsymbol{\Sigma}}\widetilde{\mathbf{w}}+\lambda \left(\mathbf{v}_+ + \mathbf{v}_-\right)'\boldsymbol{1}_N,
\end{equation}
where $$
\widetilde{\mathcal{W}}_{LS}^{\pm}(\vartheta,\mathbf{V})=\left\{(\widetilde{\mathbf{w}},\gamma,\mathbf{v}_+,\mathbf{v}_-)\in\mathbb{R}^{2N+K+1}: \mathbf{V}\widetilde{\mathbf{w}}=\mathbf{v}_+-\mathbf{v}_-,\quad [\widetilde{\mathbf{w}},\gamma]\in\widetilde{\mathcal{W}}_{LS}(\vartheta,\mathbf{V})\right\},$$ 
$\boldsymbol{\Sigma_f} = \mathbf{Cov}(\mathbf{F})$,  $\mathbf{F}$ denotes the factors from PCA, RP-PCA, IPCA models, $\widetilde{\boldsymbol{\Sigma}} = \mathbf{P}[\boldsymbol{\Lambda_f}+\lambda_2\mathbf{V'V}]\mathbf{P}'$, and $\mathbf{V}$ is an $N \times K$ mapping matrix, the eigenvector corresponding to the first $K$ largest eigenvalues of the PCA, RP-PCA, or IPCA covariance matrix; $\mathbf{v}_+, \mathbf{v}_-$ are $N \times 1$ vectors, which denote the positive and negative part of $\mathbf{V}\widetilde{\mathbf{w}}$. Importantly, the final objective function in the optimization is quadratic, and the constraints are linear. Hence, the corresponding problem falls into the general class of QP problems that we solve using our framework. In the following empirical analysis we also use \( \vartheta=0.2 \), \( r_f = 0 \), \( L_j=-0.08 \), and \( U_j=0.08 \) for all \( j=1,\ldots,N \) as in all the previous methods.

\section{Empirical Results}\label{sec:empirics}

We gather both daily and monthly data for all stocks traded on the NYSE, Amex, and Nasdaq from January 1965 to December 2022. The daily and monthly stock returns, adjusted for splits and dividends, are sourced from the Center for Research in Security Prices (CRSP). Additionally, we obtain quarterly accounting-related information for public firms from the Compustat dataset, which includes metrics such as BE (book equity), AT (total assets), and CTO (capital turnover). Following the methodologies of \citet{FAMA1993} and \citet{freyberger2020dissecting}, we merge the returns data with the firm-specific information, introducing a 6-month lag for all firms to ensure our results are genuinely out-of-sample.

After obtaining the merged datasets, we construct 33 characteristics, with a full list provided in the Appendix, using data from firms in the Compustat dataset as described by \citet{freyberger2020dissecting} and references therein. For imputation purposes, we adopt the backward cross-sectional model proposed by \citet{bryzgalova2022missing}. In our research, we utilize the stock universe defined by \citet{asness2013value}, to which we refer as the AMP universe. To assemble this universe, we implement a rolling window approach and select stocks in each window based on specific criteria. Initially, in our market capitalization-based stock selection, we exclude the smallest market capitalization stocks, focusing mainly on large- and mid-cap stocks, which together account for 90\% of the overall market capitalization. Subsequently, we filter out stocks priced below a designated threshold, ensuring the exclusion of penny stocks. Finally, to maintain the consistency of the dataset, we remove stocks with significant missing data in the last selection phase.

Depending on the specific model under consideration, we use either daily or monthly simple returns from the constructed AMP universe. For daily returns, the AMP universe typically consists of 500 to 1,000 stocks at any given time within a one-year rolling window. For monthly returns, we employ a rolling window of 20 years, resulting in an AMP universe of approximately 900 tickers for each window. Crucially, our methodology in constructing the rolling window-specific assets universe ensures that the portfolio and its performance are not affected by survivorship bias.

\begin{figure}[ht]
\centering
\resizebox*{10cm}{!}{\includegraphics{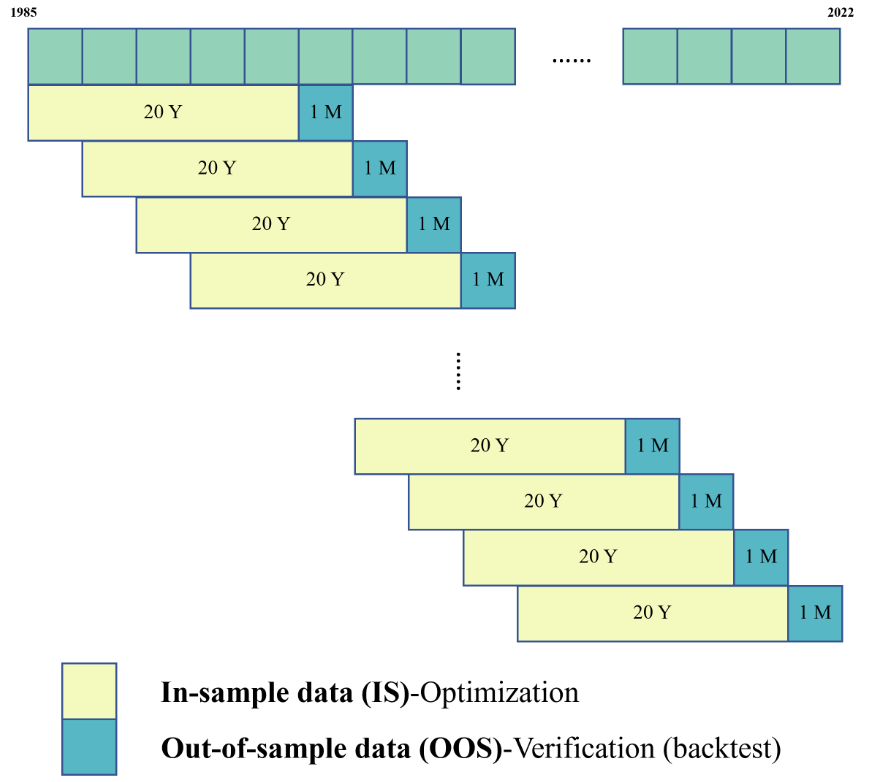}}
\caption{\protect \footnotesize Summary of the rolling window analysis. We use data going back to 1985 and slide 20 years of monthly returns to estimate the parameters, with monthly rebalancing and performance updates.} \label{fig:rollingwindow_plot}
\end{figure}

In portfolio optimization, the evaluation of out-of-sample performance of a specific model is often of interest. For this purpose, a rolling window backtest analysis is typically employed. Figure \ref{fig:rollingwindow_plot} illustrates our rolling window scheme for the monthly data utilized in factor-based models (AP-Trees and all PCA-based models). We partition the 38 years of data into a 20-year training sample (1985-2004) and allocate the subsequent 18 years (2005-2022) for out-of-sample rolling window analysis. This involves monthly reestimation of all model parameters and optimization of portfolio weights. For models investing in individual stocks without leveraging information from stock-specific factors, we adopt a rolling window of daily returns with a one-year look-back period. The rebalancing occurs monthly, commencing on the same start date as in the case of the 20-year window of monthly returns. Thus, all out-of-sample results presented in the following sections span the identical time frame and maintain consistent rebalancing frequency. In terms of portfolio weight constraints, for all the scenarios discussed, we restrict asset concentration to no more than $\pm 8\%$ for a single asset and cap short positions at $20\%$ of the total capital. We selected these thresholds to mirror a realistic industry environment, as described in \cite{lunde2016econometric}.

For our benchmark methods, we utilize daily data and incorporate three distinct mean shrinkage estimators as proposed by \citet{wang2014non} and \citet{bodnar2019optimal}. Additionally, we employ four covariance matrix estimators: the Sample Covariance Matrix, POET (as detailed by \citet{fan2013large}), and both the Ledoit \& Wolf Linear and Non-Linear Shrinkage methods from \citet{Ledoit:04} and \citet{ledoit2020analytical}, respectively, as discussed in Section \ref{sec:covariance_model}.

In contrast, we evaluate these benchmarks against the $\ell_1$+$\ell_2^2$ regularized minimum variance portfolio. The out-of-sample annualized Sharpe ratio results of these methods are illustrated in Figure \ref{fig:bench_shar}. This figure showcases heatmaps that detail the out-of-sample Sharpe ratios for portfolios, rebalanced monthly using one year of daily data for parameters. Meanwhile, Figure \ref{fig:mini_var_sharpe} highlights the minimum variance portfolio that implements a long-short constraint, complemented by $\ell_1$ and $\ell_2^2$ shrinkage. Further insights into Sharpe ratios, derived from various mean and covariance matrix estimator combinations, are provided in Figure \ref{fig:mini_l1l2}.

From a vertical perspective, the heatmap sorts portfolios based on five mean estimators: the Sample Mean, Mean Shrinkage I (from \citet{wang2014non}), Mean Shrinkage II and III (both from \citet{bodnar2019optimal}), and a minimum variance portfolio that does not factor in mean estimation. Horizontally, the heatmap is structured according to covariance matrix estimators, namely: the Sample Covariance Matrix (SCM), POET (by \citet{fan2013large}), Linear Shrinkage (L\&W-LS) from \citet{Ledoit:04}, and Nonlinear Shrinkage (L\&W-NLS) from \citet{ledoit2020analytical}.

Intriguingly, the minimum variance portfolios showcased in Panel (a) and the base of Panel (b) in Figure \ref{fig:bench_shar} outperform maximum Sharpe ratio portfolios that apply shrinkage either to the mean, the covariance matrix, or both. Moreover, the norms of the minimum-variance regularized portfolio in Panel (a) of Figure \ref{fig:bench_shar} in most of the cases mirror or surpass the performance of the covariance matrix shrinkage methods when applied to a minimum-variance portfolio. These findings indicate that the $\ell_1$ and $\ell_2^2$ regularized portfolio methods perform similarly to best performing covariance shrinkage estimators.

{\captionsetup{labelfont={color=black},textfont={color=black}}
\begin{figure}[ht]
\centering
\subfloat[$\min Var + \ell_1 + \ell_2^2$\label{fig:mini_l1l2}]{%
\resizebox*{6.5cm}{!}{\includegraphics{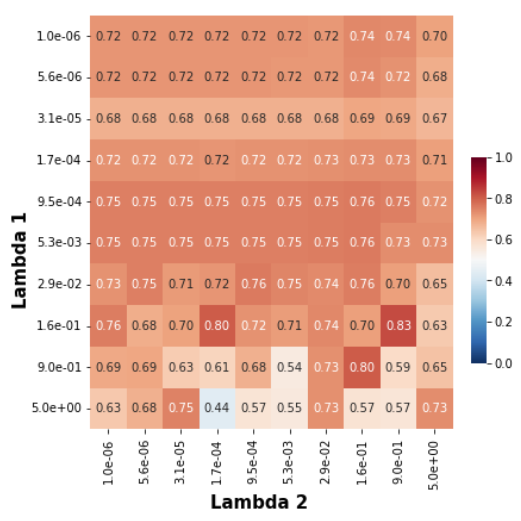}}}\hspace{5pt}
\subfloat[$\max SR$ and $\min Var$ with mean and covariance matrix shrinkage\label{fig:mini_var_sharpe}]{%
\resizebox*{6.5cm}{!}{\includegraphics{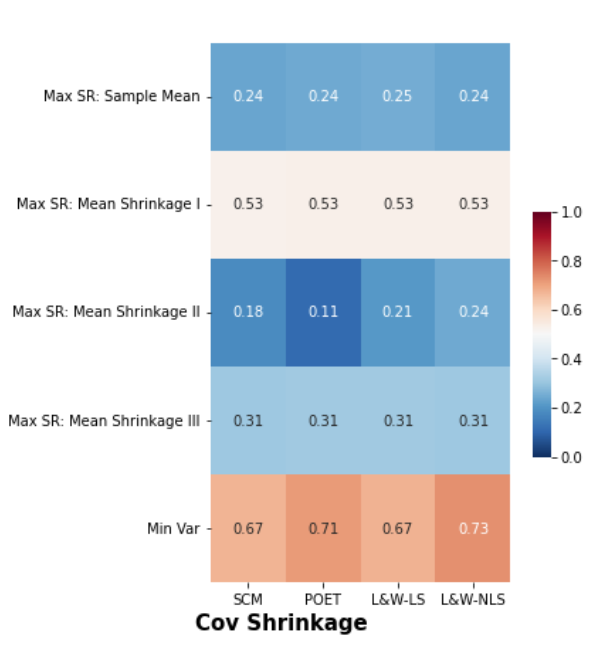}}}\hspace{5pt}
\caption{\protect \footnotesize Heatmaps of annualized Sharpe ratios for different minimum-variance portfolios on individual assets from the AMP universe using one year of daily returns data and monthly rebalancing. \textbf{Left:} The annualized Sharpe ratios of $\ell_1+\ell_2^2$ regularized minimum-variance portfolio strategy for different regularization strength parameters. \textbf{Right:} Row-wise: different mean estimators and portfolio without the dependency on the mean. Namely, maximum Sharpe ratio portfolio with four types of mean estimators: Sample Mean, Mean Shrinkage I from \citet{wang2014non}, Mean Shrinkage II from \citet{bodnar2019optimal}, Mean Shrinkage III from \citet{bodnar2019optimal}, and minimum variance that does not depend on the mean estimation. Column wise: different covariance matrix estimators: Sample Covariance Matrix (SCM), POET from \citet{fan2013large}, Linear Shrinkage covariance matrix estimator from \citet{Ledoit:04} (L\&W-LS), Nonlinear Shrinkage covariance matrix estimator from \citet{ledoit2020analytical} (L\&W-NLS).} \label{fig:bench_shar}
\end{figure}}

The findings presented in Figure \ref{fig:bench_shar}b show that the minimum-variance portfolio consistently outperforms the maximum Sharpe ratio portfolio, regardless of the shrinkage applied. This observation aligns with our earlier comments regarding the inherent noisiness of individual stock means. Optimization strategies based on individual stocks frequently yield suboptimal out-of-sample results. Subsequent analyses will highlight that managed portfolios can mitigate the idiosyncratic noise present in individual stock returns, thereby delivering optimal portfolios with superior out-of-sample performance.

Figure \ref{fig:tree_heatmap} displays the out-of-sample annualized Sharpe ratios for AP-Trees portfolios, which are rebalanced monthly. These portfolios are derived from the $\ell_1+\ell_2^2$ regularized maximum Sharpe ratio portfolio strategy, as outlined in \eqref{eq:ela_factor_modified}. Each heatmap represents a unique managed portfolio, distinguished by market capitalization and paired with two other characteristics from Table \ref{table:characteristics}. The differences across heatmaps also reflect variations in the regularization strength parameters, $\lambda_1$ and $\lambda_2$. In all cases, a 20-year rolling window of monthly data is used. The short-selling constraint is set at $\vartheta=0.2$, and the maximum concentration in an individual managed portfolio is capped at $8\%$. We observe a notable improvement in Sharpe ratios compared to the top-performing portfolios invested in individual assets. This suggests that grouping stocks with analogous characteristics into managed portfolios effectively diminishes noise and enhances mean prediction.

Next, we examine the three PCA-based models outlined in Section \ref{sec:covariance_model}. Figure \ref{fig:heatmap_benchmark} presents heatmaps depicting the out-of-sample Sharpe ratios for a monthly rebalanced portfolio that invests in $K=2,\ldots,6$ factors from the PCA, RP-PCA, and IPCA models, respectively. The figure comprises 15 heatmaps, all on a consistent scale. Each heatmap demonstrates performance across different levels of $\ell_1$ and $\ell_2^2$ regularization parameters, taken from an exponential grid spanning $\lambda_1=10^{-6},\ldots,5$ and $\lambda_2 = 10^{-6},\ldots,5$. Empirically, within these parameter ranges, the regularization has the most pronounced impact on the portfolio weights across all models. For every model and every factor count $K$, the proposed regularization consistently enhances performance. The peak performance is observed with $K=6$ factors. Specifically, the Sharpe ratios rise for (i) the PCA from 1.52 to 2.00; (ii) the RP-PCA  from 2.12 to 3.40; and (iii) the IPCA from 3.75 to 4.93. Furthermore, as illustrated in Figure \ref{fig:heatmap_benchmark}, there is a marked improvement as the number of components from PCA and IPCA increases. Exploring a broad range of regularization parameters enables us to pinpoint their most effective values. For \( \lambda_1 \), the optimal value is approximately \( 1.7 \times 10^{-4} \), while for \( \lambda_2 \), it lies between \( 1.0 \times 10^{-6} \) and \( 2.9 \times 10^{-2} \). Across all values of $K$ and various regularization strengths, the RP-PCA model consistently surpasses the corresponding PCA models. The most outstanding performer among all considered models is the IPCA model with 6 factors and combined $\ell_1$ and $\ell_2^2$ shrinkage.

{
\captionsetup{labelfont={color=black},textfont={color=black}}
\begin{figure}
\centering
\resizebox*{14.5cm}{!}{\includegraphics{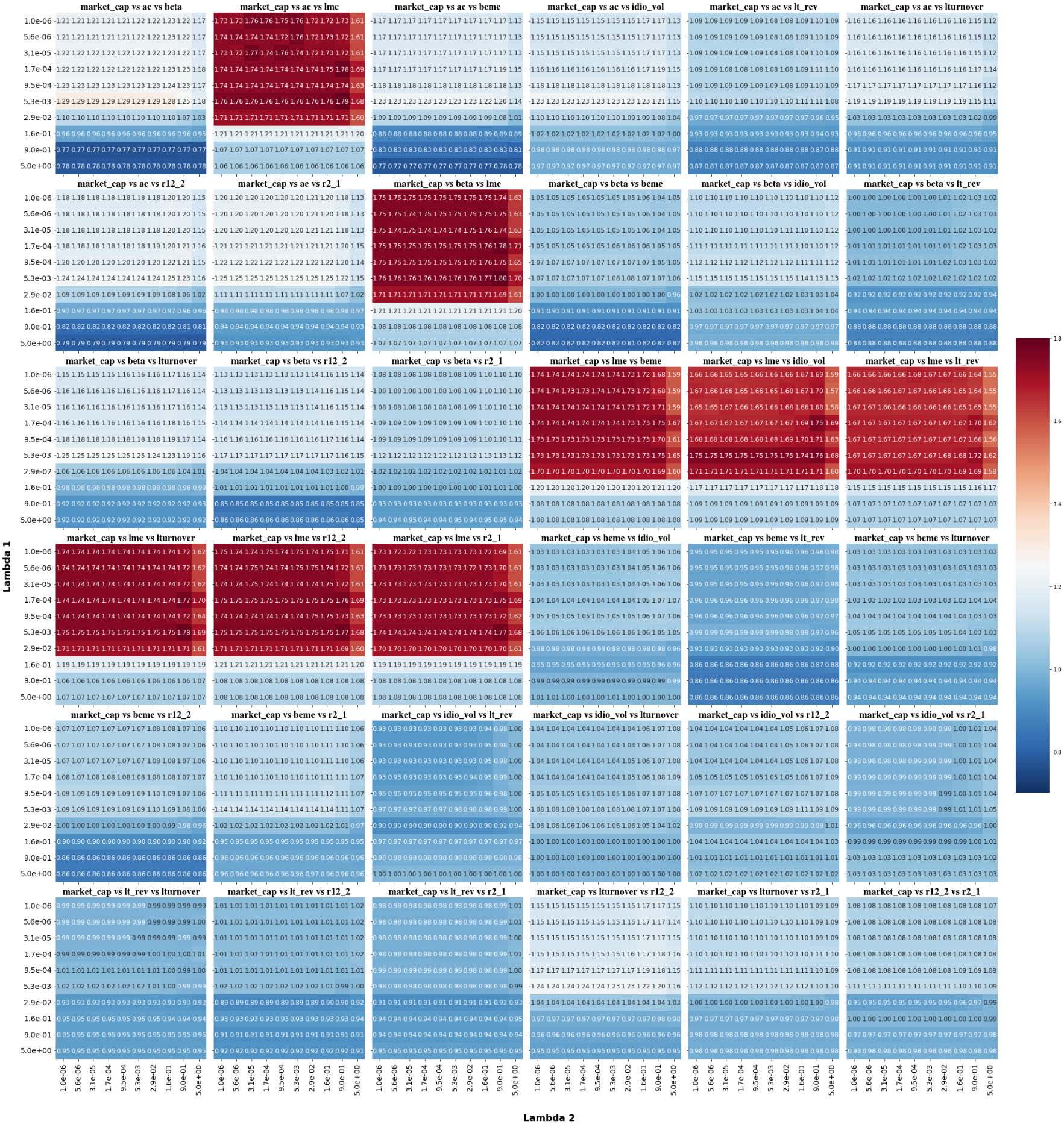}}
\caption{\protect \footnotesize Thirty-six heatmaps of out-of-sample annualized Sharpe ratios from monthly rebalanced AP-Trees portfolios obtained from $\ell_1+\ell_2^2$ regularized maximum Sharpe ratio portfolio strategy computed using \eqref{eq:MaximumSharperatio_model}. Different heatmaps correspond to  different managed portfolios of market capitalization with the combination of another two characteristics from Table \ref{table:characteristics}, and different regularization strength parameters, $\lambda_1$ and $\lambda_2$.
In all the cases, we use a rolling window of 20 years of monthly data with short-selling constraint $\vartheta=0.2$ and no additional constraints.} \label{fig:tree_heatmap}
\end{figure}}

{
\captionsetup{labelfont={color=black},textfont={color=black}}
\begin{figure}
\centering
\resizebox*{14.5cm}{!}{\includegraphics{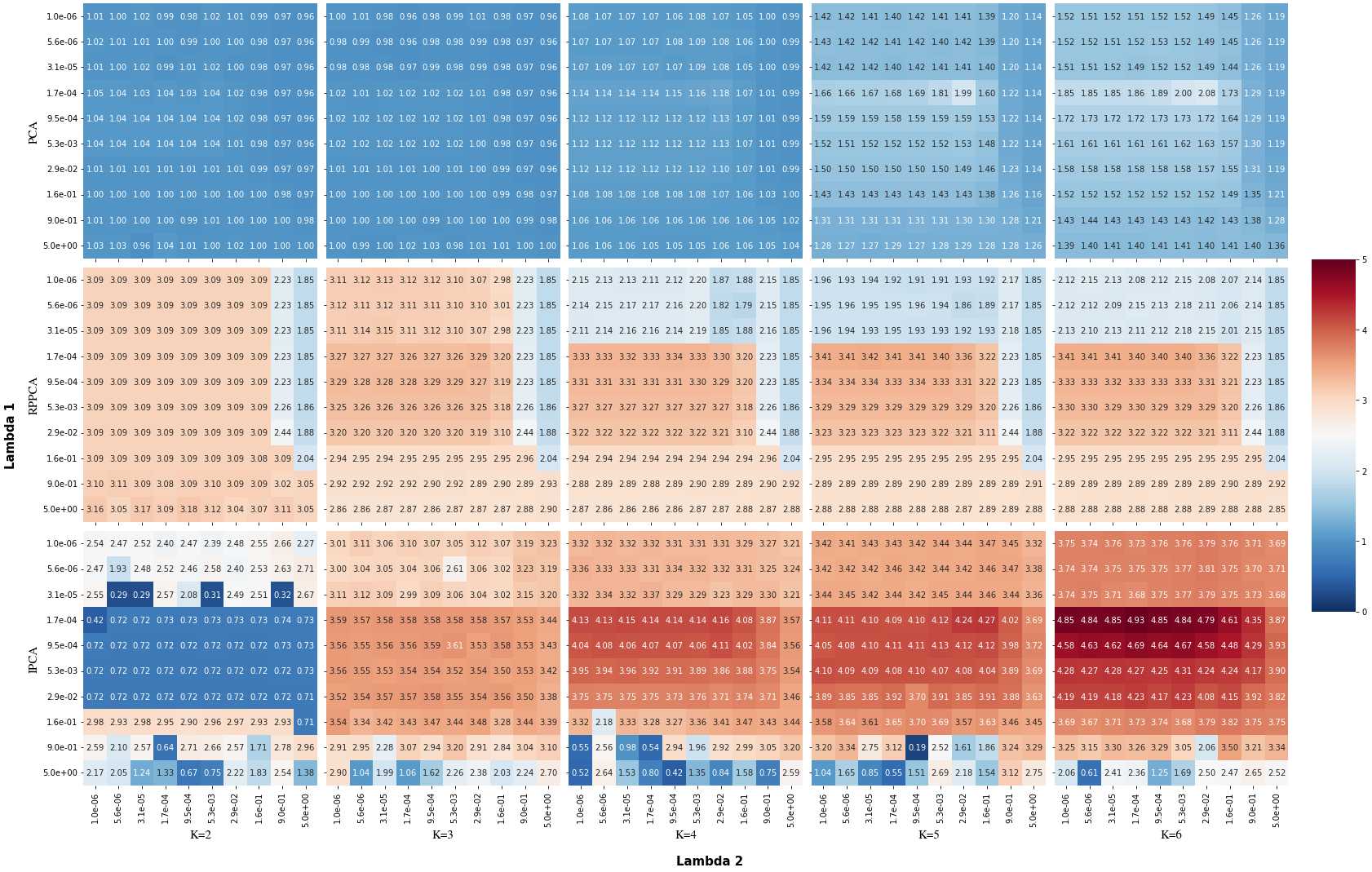}}
\caption{\protect \footnotesize Fifteen heatmaps of out-of-sample performance gains from monthly rebalanced portfolio in the annualized Sharpe ratios of $\ell_1+\ell_2^2$ regularized maximum Sharpe ratio portfolio strategy. The strategy varies based on the number of components from PCA, RP-PCA, and IPCA estimated covariance matrix, and different regularization strength parameters, $\lambda_1$ and $\lambda_2$.
In all the cases, we use the maximum Sharpe ratio portfolio estimated based on the last 20 years of data with short-selling constraint $\vartheta=0.2$ and no additional constraints.
\textbf{Columns}: Different size components from PCA, RP-PCA, IPCA models.
\textbf{First Row}:  Annualized Sharpe ratios for the covariance matrix derived from PCA factors, regularized with $\ell_1+\ell_2^2$.
\textbf{Second Row}:  Annualized Sharpe ratios for the covariance matrix derived from RP-PCA factors, regularized with $\ell_1+\ell_2^2$.
\textbf{Third Row}:  Annualized Sharpe ratios for the covariance matrix derived from IPCA factors, regularized with $\ell_1+\ell_2^2$.} \label{fig:heatmap_benchmark}
\end{figure}}

{
\captionsetup{labelfont={color=black},textfont={color=black}}

\begin{figure}
\centering

\subfloat[Underwater: PCA \& $\max SR$\label{fig:underwater1_pca}]{%
\resizebox*{6.5cm}{!}{\includegraphics{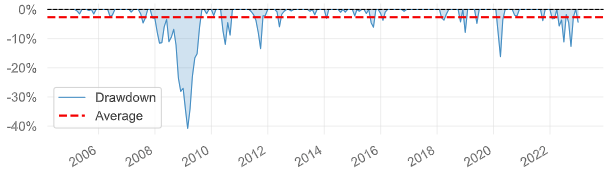}}}\hspace{5pt}
\subfloat[Underwater: PCA \& $\max SR+\ell_1+\ell_2^2$\label{fig:underwater2_pca}]{%
\resizebox*{6.5cm}{!}{\includegraphics{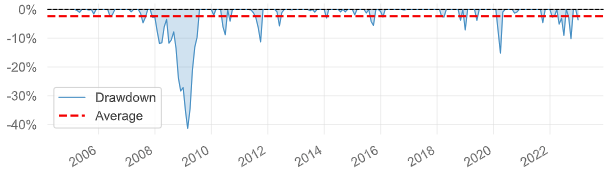}}}\hspace{5pt}

\subfloat[Monthly Ret.: PCA \& $\max SR$\label{fig:monthly_ret1_pca}]{%
\resizebox*{6.5cm}{!}{\includegraphics{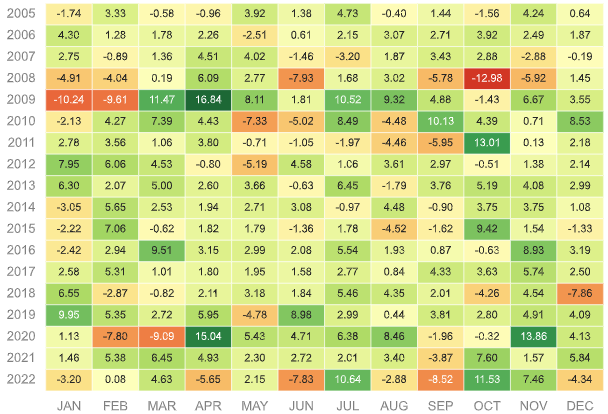}}}\hspace{5pt}
\subfloat[Monthly Ret.: PCA \& $\max SR+\ell_1+\ell_2^2$\label{fig:monthly_ret2_pca}]{%
\resizebox*{6.5cm}{!}{\includegraphics{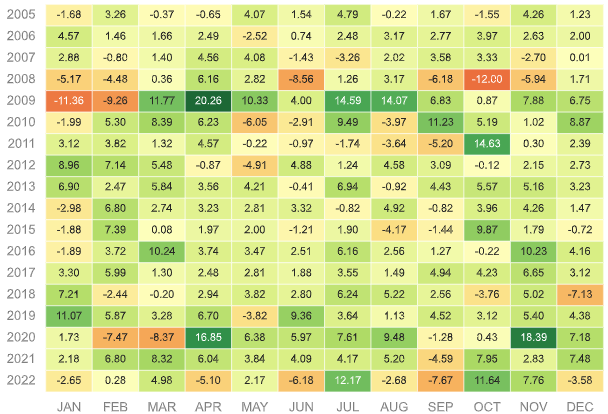}}}\hspace{5pt}

\caption{\protect \footnotesize Out-of-sample underwater plots (Panels (a) and (b)) and Monthly Returns performance (Panels (c) and (d)) for our long-short maximum-Sharpe ratio with PCA factors without regularization (Panels (a) and (c)), and with $\ell_1$+$\ell_2^2$ regularization from \eqref{eq:ela_factor_modified} (Panels (b) and (d)).}

\label{fig:monthly_ret_comp_pca}
\end{figure}
}

{
\captionsetup{labelfont={color=black},textfont={color=black}}
\begin{figure}
\centering

\subfloat[Underwater: RP-PCA \& $\max SR$\label{fig:underwater1_rppca}]{%
\resizebox*{6.5cm}{!}{\includegraphics{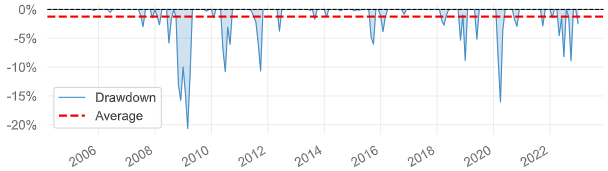}}}\hspace{5pt}
\subfloat[Underwater: RP-PCA \& $\max SR+\ell_1+\ell_2^2$\label{fig:underwater2_rppca}]{%
\resizebox*{6.5cm}{!}{\includegraphics{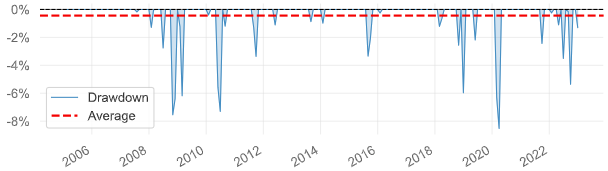}}}\hspace{5pt}

\subfloat[Monthly Ret.: RP-PCA \& $\max SR$\label{fig:monthly_ret1_rppca}]{%
\resizebox*{6.5cm}{!}{\includegraphics{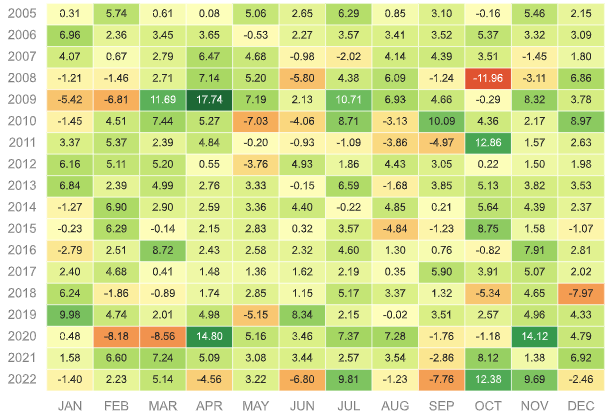}}}\hspace{5pt}
\subfloat[Monthly Ret.: RP-PCA \& $\max SR+\ell_1+\ell_2^2$\label{fig:monthly_ret2_rppca}]{%
\resizebox*{6.5cm}{!}{\includegraphics{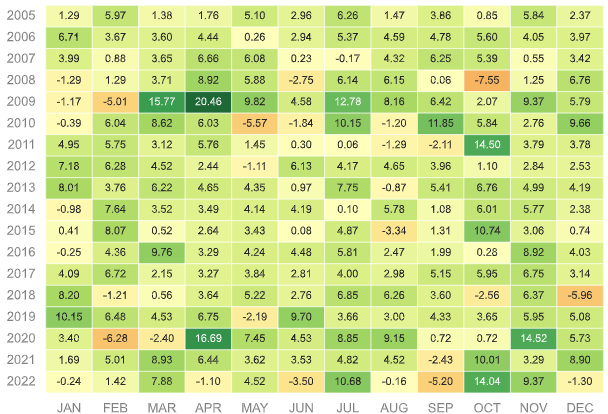}}}\hspace{5pt}
\caption{\protect \footnotesize Analogous to Figure \ref{fig:monthly_ret_comp_pca} but for the RP-PCA model.}
\label{fig:monthly_ret_comp_rppca}
\end{figure}
}

{
\captionsetup{labelfont={color=black},textfont={color=black}}
\begin{figure}
\centering
\subfloat[Underwater: IPCA \& $\max SR$\label{fig:underwater1}]{%
\resizebox*{6.5cm}{!}{\includegraphics{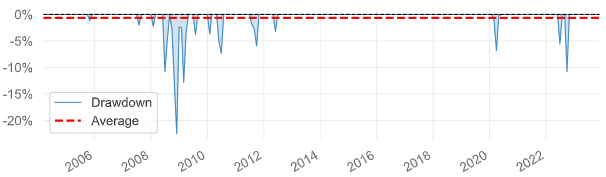}}}\hspace{5pt}
\subfloat[Underwater: IPCA \& $\max SR+\ell_1+\ell_2^2$\label{fig:underwater2}]{%
\resizebox*{6.5cm}{!}{\includegraphics{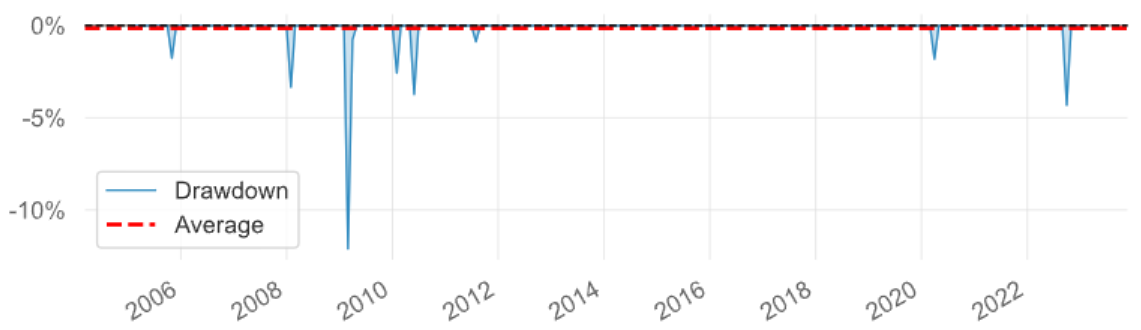}}}\hspace{5pt}
\subfloat[Monthly Ret.: IPCA \& $\max SR$\label{fig:monthly_ret1}]{%
\resizebox*{6.5cm}{!}{\includegraphics{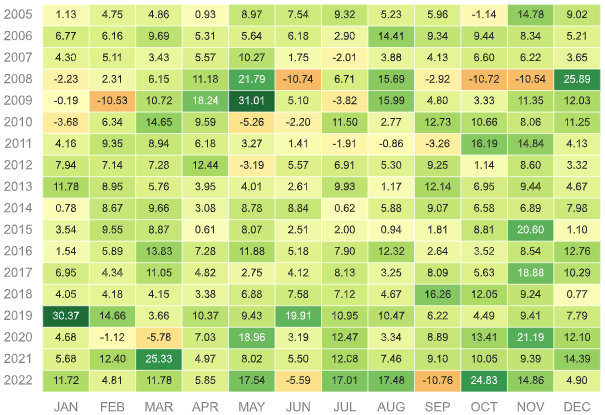}}}\hspace{5pt}
\subfloat[Monthly Ret.: IPCA \& $\max SR+\ell_1+\ell_2^2$\label{fig:monthly_ret2}]{%
\resizebox*{6.5cm}{!}{\includegraphics{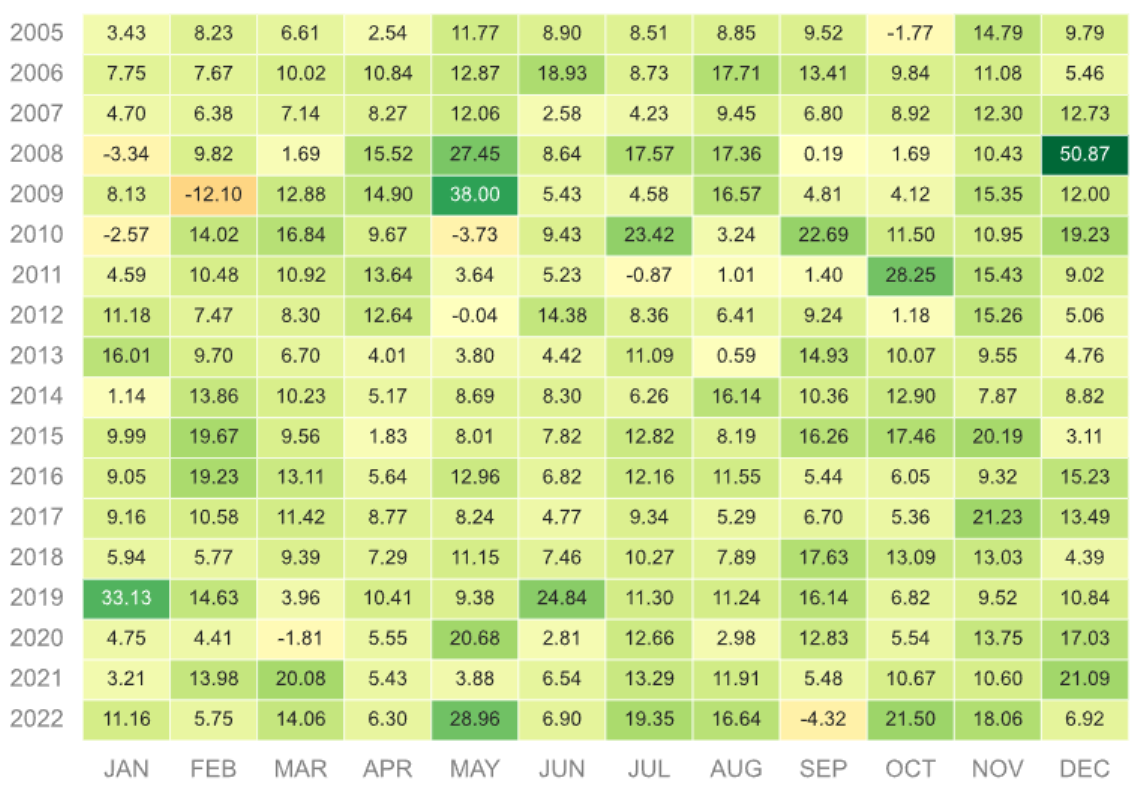}}}\hspace{5pt}
\caption{\protect \footnotesize Analogous to Figure \ref{fig:monthly_ret_comp_pca} but for the IPCA model.}
\label{fig:monthly_ret_comp}
\end{figure}
}

Figure~\ref{fig:monthly_ret_comp_pca} contrasts the performance of the PCA factor model for the maximum Sharpe ratio portfolio ($K=6$) with and without regularization on the portfolio norms, as delineated in \eqref{eq:maximumSharperatio_used} and \eqref{eq:ela_factor_modified}, respectively. Panels \ref{fig:underwater1_pca} and \ref{fig:monthly_ret1_pca} present the results for the PCA max Sharpe ratio portfolio without regularization. In contrast, panels \ref{fig:underwater2_pca} and \ref{fig:monthly_ret2_pca} showcase the results with the inclusion of $\ell_1+\ell_2^2$ regularization, utilizing the optimal $\lambda_1$ and $\lambda_2$ parameters. Both panels \ref{fig:underwater1_pca} and \ref{fig:underwater2_pca} suffer from a large  drawdown during the financial crisis. Nevertheless, in other periods, the regularized PCA factor model demonstrates enhanced performance. This distinction becomes even more evident in Panels \ref{fig:monthly_ret1_pca} and \ref{fig:monthly_ret2_pca}, where the regularized portfolio model outperforms in the majority of the months considered.

Figure~\ref{fig:monthly_ret_comp_rppca} parallels Figure~\ref{fig:monthly_ret_comp_pca} but focuses on the RP-PCA model ($K=6$). We examine both the inclusion and exclusion of the $\ell_1+\ell_2^2$ regularization, specifically selecting the optimal $\lambda_1$ and $\lambda_2$ parameters. Panels \ref{fig:underwater1_rppca} and \ref{fig:monthly_ret1_rppca} depict the underwater and monthly return plots for the RP-PCA factor model without the $\ell_1+\ell_2^2$ constraints. Conversely, panels \ref{fig:underwater2_rppca} and \ref{fig:monthly_ret2_rppca} showcase these plots with the $\ell_1+\ell_2^2$ regularization applied. The regularized RP-PCA displays a trend akin to the benchmark model (RP-PCA factor model without $\ell_1+\ell_2^2$ regularization). Notably, the application of regularization in the RP-PCA model considerably mitigates drawdowns; for instance, the maximum monthly drawdown shrinks from $-20\%$ to $-8.2\%$. This enhancement is further confirmed by panels \ref{fig:monthly_ret1_rppca} and \ref{fig:monthly_ret2_rppca}, which consistently indicate elevated returns for the regularized RP-PCA model.

Finally, Figure \ref{fig:monthly_ret_comp} offers a similar comparison but focuses on the IPCA model. Analogous to previous observations, the IPCA model augmented with the $\ell_1+\ell_2^2$ regularization for the maximum Sharpe ratio portfolio exhibits consistently fewer and smaller drawdowns compared to the unregularized IPCA model. Moreover, the monthly returns for the regularized maximum Sharpe ratio portfolio are consistently higher throughout the entire out-of-sample analysis period.

In summary, both the AP-Trees and the three PCA-based models gain significantly from the proposed regularization of the linear transformation of the portfolio norms, $\mathbf{V}\mathbf{w}$. Among all the models considered, the IPCA model stands out as the top performer. The out-of-sample performance of both the original and regularized IPCA models is truly exceptional. Even when accounting for market frictions, such as transaction costs, implementation lags, liquidity concerns, and potential complications arising from the construction of certain asset-specific characteristics, a significant portion of this remarkable performance is expected to remain intact. While various methods exist to further refine the investment process and mitigate the effects of these market frictions, delving into them remains a topic for future research. It is worth noting that our IPCA model results without regularization are in agreement with findings from the original IPCA paper (refer to the Sharpe ratios of the tangent portfolios in Table 5 of \citealp{KELLY2019501}). Our regularization of the linear combinations of the portfolio norms further enhances the model's efficacy. Moreover, the portfolio results presented in this study deviate from the original \citet{KELLY2019501} portfolio due to the incorporation of a 20\% long-short constraint, a cap of 8\% on individual positions, and trading restrictions to only the AMP universe of mid- and large-capitalization stocks. That the portfolio, despite these constraints, can achieve such impressive monthly returns, high annualized Sharpe ratios, and minimal drawdowns over nearly 18 years of out-of-sample rolling windows is intriguing and noteworthy.

Table~\ref{table:quanstat_table} presents key performance metrics across various benchmarks: the S\&P 500 index; two minimum variance portfolios utilizing Ledoit \& Wolf's linear and non-linear shrinkage covariance matrices; the best performing AP-Trees and factor portfolios based on PCA, RP-PCA, and IPCA with $K=6$ models and regularization. The latter is also presented without regularization, as per the original model by \citet{KELLY2019501}. The initial three benchmarks are based on individual daily stock returns. For the AP-Trees, we employ 360 managed portfolios conditionally sorted based on size, beta, and lagged market capitalization using a depth-three tree (refer to Figure \ref{fig:tree_heatmap} for the highest Sharpe). The PCA and RP-PCA models utilize 330 single-sort monthly managed portfolios. Contrarily, the IPCA model strictly operates on individual stock returns, incorporating stock-specific firm data. The first IPCA portfolio is the constrained tangency portfolio without regularization, with its covariance matrix determined via the IPCA factors model. The subsequent IPCA portfolio is the same but with optimal regularization employed. In summation, the regularized IPCA portfolio, results in an annualized Sharpe ratio of 4.91, and it surpasses all other methods across nearly every metric considered. The regularized RP-PCA is performs best in terms of the lowest maximum drawdown, highest information ratio, and the lowest loss in the worst month. It also has lower volatility than the IPCA models.

{
\captionsetup{labelfont={color=black},textfont={color=black}}
\begin{table}
\caption{\protect
\footnotesize Key performance metrics from a rolling window exercise with monthly rebalancing from 2005-01-31 until 2022-12-31. \textbf{First Column:} S\&P500 index as a long-only market benchmark. \textbf{Second Column}: The minimum variance optimal portfolio with sample covariance matrix computed from one year look-back window of daily returns, and long-short constrain as in \eqref{eq:l1meanvariance_old}. \textbf{Third Column}: The minimum variance optimal portfolio with \citet{ledoit2020analytical} nonlinear shrinkage covariance matrix computed from one year look-back window of daily returns, and long-short constraints. \textbf{Fourth Column}: 20 years of look-back window of AP-Trees managed portfolio monthly returns using maximum Sharpe ratio optimal portfolio strategies as in \eqref{ep:maximumShaperation_final} with long-short constraints, $\ell_1$ and $\ell^2_2$ shrinkage. Remaining columns use 20 years of look-back window of monthly returns and different maximum Sharpe ratio optimal portfolio strategies with long-short constraints as in \eqref{eq:MaximumSharperatio_model} and the covariance matrix of factor portfolios for $K=6$ estimated via \textbf{Fifth Column}: PCA model with $\ell_1$ and $\ell^2_2$ regularized maximum Sharpe ratio portfolio; \textbf{Sixth Column}: RP-PCA model with $\ell_1$ and $\ell_2$ regularized maximum Sharpe ratio portfolio; \textbf{Seventh Column}: IPCA model with maximum Sharpe ratio portfolio without $\ell_1$ and $\ell^2_2$ regularization. \textbf{Eighth Column}: IPCA model with $\ell_1$ and $\ell^2_2$ regularized maximum Sharpe ratio portfolio.}
\resizebox{\textwidth}{!}{%
\begin{tabular}{lllllllll}
\toprule
{} &     Market & Sample Cov &  L\&W-NLS &   AP-Trees  &      PCA &      RP-PCA & IPCA &    IPCA\\
{} &     S\&P 500 & $\min Var$ & $\min Var$& $\max SR+\ell_1+\ell_2^2$& $\max SR+\ell_1+\ell_2^2$ &  $\max SR+\ell_1+\ell_2^2$ & $\max SR$ &$\max SR+\ell_1+\ell_2^2$ \\
\midrule
Start Period              &  2005-02-02 &         2005-02-02 &            2005-02-02 &      2005-01-31 &  2005-01-31 &  2005-01-31 &       2005-01-31 &           2005-01-31 \\
End Period                &  2022-12-30 &         2022-12-30 &            2022-12-30 &      2022-12-31 &  2022-12-31 &  2022-12-31 &       2022-12-31 &           2022-12-31 \\
Risk-Free Rate            &           0 &                  0 &                     0 &               0 &           0 &           0 &                0 &                    0 \\
Time in Market            &        1.0 &                1.0 &                   1.0 &             1.0 &         1.0 &         1.0 &              1.0 &                  1.0 \\
CAGR\%                     &        0.07 &               0.08 &                  0.08 &            0.31 &        0.35 &        0.59 &             1.24 &                  \textbf{2.1} \\
Sharpe                    &         0.5 &               0.67 &                  0.73 &             1.8 &        1.85 &         3.4 &             3.74 &                 \textbf{4.91} \\
Smart Sharpe              &        0.45 &               0.61 &                  0.66 &            1.75 &        1.67 &        3.29 &             3.73 &                 \textbf{4.69} \\
Omega                     &        1.15 &               1.14 &                  1.15 &            4.09 &        3.96 &       13.17 &            16.62 &                \textbf{72.24} \\
Max Drawdown              &       -0.52 &              -0.35 &                 -0.38 &           -0.18 &       -0.41 &       \textbf{-0.09} &            -0.22 &                -0.12 \\
Longest DD Days           &        1947 &               1396 &                  1639 &             212 &         609 &         151 &              212 &                   \textbf{61} \\
Volatility (ann.)         &        0.15 &               \textbf{0.12} &                  \textbf{0.12} &            0.16 &        0.17 &        0.14 &             0.23 &                 0.25 \\
R\textasciicircum 2                       &           0 &                  0 &                     0 &             0.8 &        \textbf{0.91} &        0.88 &             0.39 &                 0.24 \\
Information Ratio         &           0 &                  0 &                     0 &            0.84 &        1.32 &        \textbf{2.24} &             1.25 &                 1.51 \\
Calmar                    &        0.13 &               0.22 &                  0.22 &             1.7 &        0.86 &        6.96 &             5.52 &                \textbf{17.39} \\
Skew                      &       -0.71 &              -0.54 &                 -0.75 &            0.73 &        0.12 &        0.34 &             0.36 &                 \textbf{1.37} \\
Kurtosis                  &        77.1 &               17.4 &                 17.64 &            3.77 &        \textbf{1.23} &        1.58 &             1.89 &                 5.99 \\
Expected Yearly           &        0.07 &               0.08 &                  0.08 &            0.31 &        0.35 &        0.59 &             1.23 &                 \textbf{2.09} \\
Kelly Criterion           &        0.27 &               0.18 &                  0.17 &            0.57 &        0.57 &        0.81 &             0.85 &                 \textbf{0.95} \\
Max Consecutive Wins      &         1.0 &               12.0 &                  12.0 &            14.0 &        15.0 &        30.0 &             92.0 &                 \textbf{93.0} \\
Max Consecutive Losses    &         \textbf{1.0} &                9.0 &                  11.0 &             4.0 &         5.0 &         2.0 &              3.0 &                  \textbf{1.0} \\
Best Month                &        0.13 &               0.11 &                  0.12 &            0.26 &         0.2 &         0.2 &             0.31 &                 \textbf{0.51} \\
Worst Month               &       -0.17 &              -0.17 &                 -0.17 &           -0.09 &       -0.12 &       \textbf{-0.08} &            -0.11 &                -0.12 \\
Best Year                 &         0.3 &               0.18 &                  0.21 &            1.15 &        1.01 &        1.31 &              2.6 &                 \textbf{3.46} \\
Worst Year                &       -0.38 &              -0.19 &                 -0.19 &            0.06 &       -0.25 &        0.31 &             0.55 &                 \textbf{1.39} \\
Avg. Drawdown             &       -0.07 &              \textbf{-0.02} &                 \textbf{-0.02} &           -0.04 &       -0.05 &       -0.03 &            -0.07 &                -0.03 \\
Avg. Drawdown Days        &         166 &                 30 &                    \textbf{27} &              55 &          68 &          43 &               58 &                   33 \\
Beta                      &           - &              -0.02 &                 -0.02 &            0.92 &        \textbf{1.08} &        0.87 &             0.93 &                 0.79 \\
Alpha                     &           - &               0.08 &                  0.09 &            0.21 &        0.24 &        0.42 &             0.78 &                 \textbf{1.15} \\
Correlation               &           - &             -0.03\% &                -0.03\% &           0.89\% &       \textbf{0.95\%} &       0.94\% &            0.62\% &                0.49\% \\

\bottomrule
\end{tabular}
}
\label{table:quanstat_table}
\end{table}}

{
\captionsetup{labelfont={color=black},textfont={color=black}}
\begin{figure}
\centering
\subfloat[EOY Returns\label{fig:EOY_return}]{%
\resizebox*{6.5cm}{!}{\includegraphics{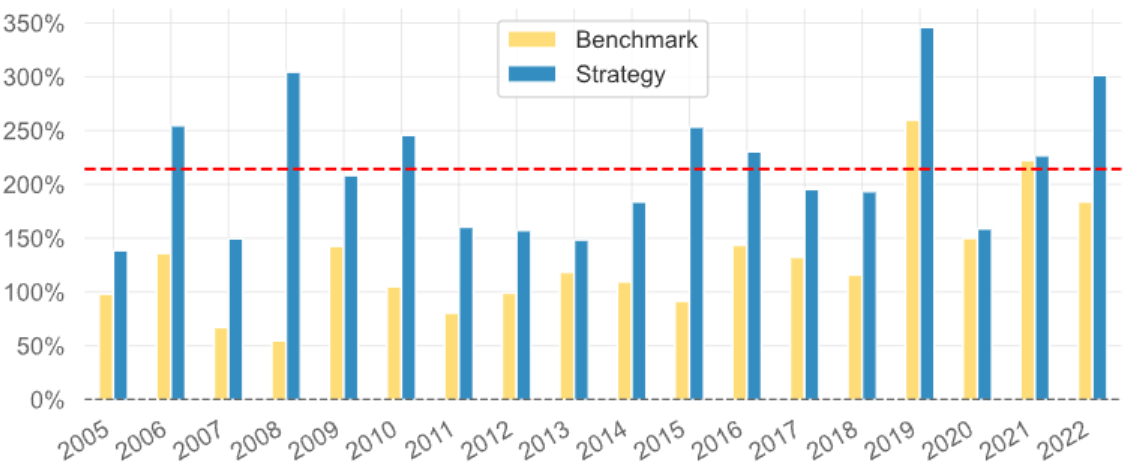}}}\hspace{5pt}
\subfloat[Distribution of Monthly Returns\label{fig:Dist_monthly_ret}]{%
\resizebox*{6.5cm}{!}{\includegraphics{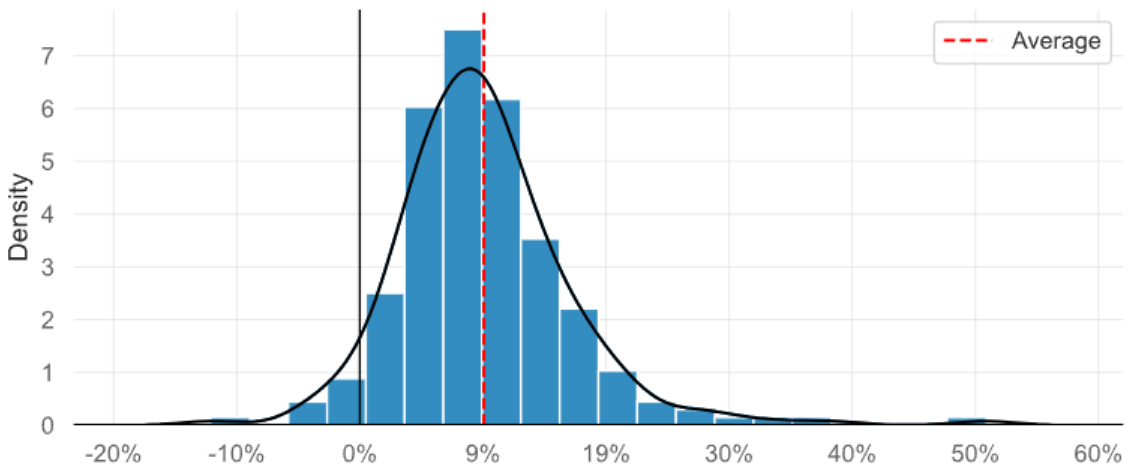}}}\hspace{5pt}
\subfloat[Rolling Sharpe (6M)\label{fig:rolling_sharpe}]{%
\resizebox*{6.5cm}{!}{\includegraphics{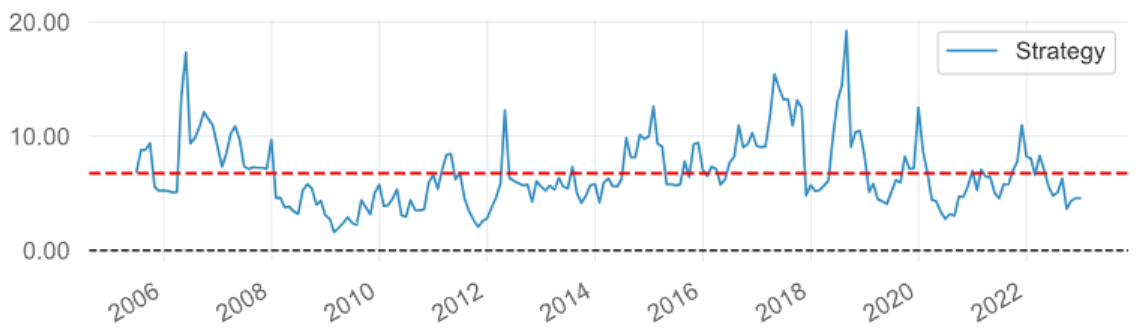}}}\hspace{5pt}
\subfloat[Rolling Beta (6M\&12M)\label{fig:rolling_beta}]{%
\resizebox*{6.5cm}{!}{\includegraphics{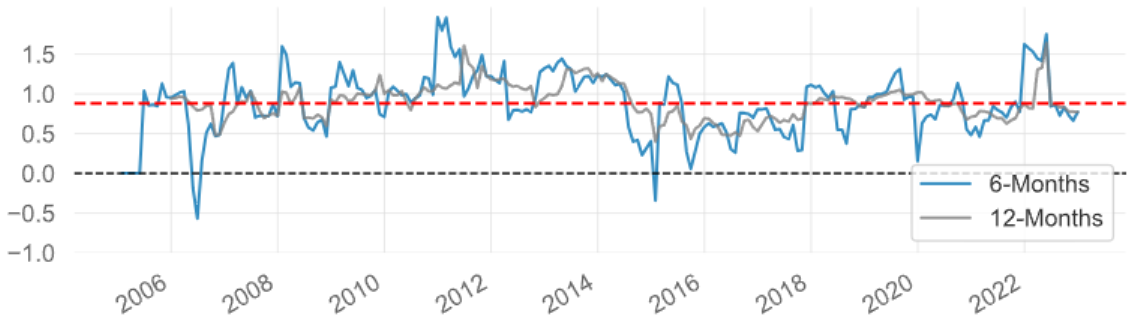}}}\hspace{5pt}
\subfloat[Rolling Volatility (6M)\label{fig:rooling_vol}]{%
\resizebox*{6.5cm}{!}{\includegraphics{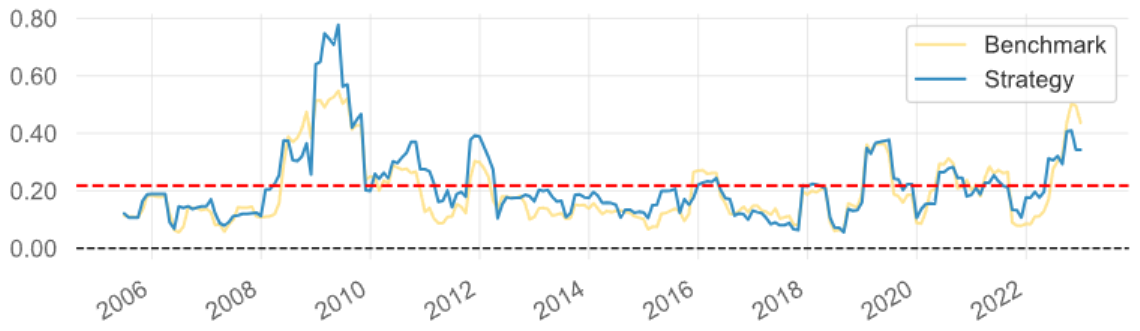}}}\hspace{5pt}
\caption{\protect \footnotesize Five plots of different out-of-sample performance measures for our long-short maximum-Sharpe ratio with $\ell_1$+$\ell_2^2$ regularization and IPCA covariance matrix estimator portfolio strategy from \eqref{eq:MaximumSharperatio_model} versus maximum-Sharpe ratio with IPCA factors covariance matrix without shrinkage as a benchmark.} 
\label{fig:quanstat_performance}
\end{figure}}

Other key performance, such as rolling beta, rolling Sharpe, and rolling volatility, of our top-performing IPCA factor model employing the maximum-Sharpe ratio portfolio with $\ell_1+\ell_2^2$ regularization, are illustrated in Figure \ref{fig:quanstat_performance}. Figure \ref{fig:quanstat_performance}a compares the annual returns of the IPCA model in maximum Sharpe ratio portfolio without (Benchmark) and with our $\ell_1+\ell_2^2$ regularization (Strategy). The regularization systematically improves the performance. Hence, it should be also simple  to calibrate the $\lambda_1$ and $\lambda_2$ parameters based on past performance. The distribution of the monthly returns is centered around $9\%$ per month, rolling 6M Sharpe ratio is very high, rolling beta (against the non-regularized benchmark) is oscillating around 1, and the rolling volatility is around 20\% only with a large burst during the Great Financial Crises and recent Covid period---variance levels deemed acceptable by quantitative portfolio managers without necessitating additional (de-)leveraging.

\section{Concluding Remarks}\label{sec:conclusions}

This study presents a unified framework for portfolio optimization using quadratic programming. This framework integrates various conventional objectives for portfolio optimization, constraints, and regularizations frequently adopted in practice. As a result, it is exceptionally suited for rapid backtesting of extensive portfolio scenarios, ensuring both accuracy and computational speed.

Employing this framework, we introduce a novel maximum Sharpe ratio portfolio problem, incorporating new types of regularizations on the norms of portfolio weights or their linear transformations. We demonstrate that, within the framework of recent tree-based and PCA-based factor models, our proposed regularization and optimization framework yield systematically enhanced returns, diminished drawdowns, reduced volatilities, and elevated Sharpe ratios for the optimal portfolios. Among the models assessed, the IPCA factor model detailed in \citet{KELLY2019501} emerges as the superior performer, especially when utilizing the proposed regularization.

In future studies, it would be intresting to delve deeper into the ramifications of transaction costs on our optimal portfolios. Factor based models because of its conditional mean prediction that depends on stock specific factors lead to inherently higher turnover numbers in portfolio optimization. Nevertheless, we believe that because of the monthly rebalancing considered in this paper, the majority of the qualitiative results will remain, also the additional smoothing $\ell_1$ constraints on the level of changes in the individual assets (similar to our $\ell_1$ regularizations) should help to reduce turnover without large impact on the performance. Additionally, integrating alternative portfolio problems within our expansive framework could help mitigate these transaction costs. As optimal portfolio weights are deduced from the inverse covariance matrix, it's vital to consider applying shrinkage methods to this matrix, which could further bolster the robustness and efficiency of the portfolio optimization. Such strategies have been explored in \citet{kourtis2012parameter}, \citet{wang2015shrinkage}, and \citet{bodnar2016direct}. Further, a comprehensive examination of the asset-specific factors in the IPCA model that significantly influence return predictions and boost portfolio performance is essential. Ideally, emphasizing factor sparsity would enhance the model's signal-to-noise ratio. This can also be achieved by incorporating sparse PCA extensions into the IPCA model.

\newpage 
\bibliographystyle{chicago}
\bibliography{main}

\begin{thebibliography}{}

\bibitem[\protect\citeauthoryear{Asness, Moskowitz, and Pedersen}{Asness
  et~al.}{2013}]{asness2013value}
Asness, C.~S., T.~J. Moskowitz, and L.~H. Pedersen (2013).
\newblock Value and momentum everywhere.
\newblock {\em The journal of finance\/}~{\em 68\/}(3), 929--985.

\bibitem[\protect\citeauthoryear{Bai and Liao}{Bai and
  Liao}{2016}]{bai2016efficient}
Bai, J. and Y.~Liao (2016).
\newblock Efficient estimation of approximate factor models via penalized
  maximum likelihood.
\newblock {\em Journal of Econometrics\/}~{\em 191\/}(1), 1--18.

\bibitem[\protect\citeauthoryear{Bai and Ng}{Bai and
  Ng}{2002}]{bai2002determining}
Bai, J. and S.~Ng (2002).
\newblock Determining the number of factors in approximate factor models.
\newblock {\em Econometrica\/}~{\em 70\/}(1), 191--221.

\bibitem[\protect\citeauthoryear{Bai and Ng}{Bai and
  Ng}{2013}]{bai2013principal}
Bai, J. and S.~Ng (2013).
\newblock Principal components estimation and identification of static factors.
\newblock {\em Journal of econometrics\/}~{\em 176\/}(1), 18--29.

\bibitem[\protect\citeauthoryear{Bodnar, Gupta, and Parolya}{Bodnar
  et~al.}{2016}]{bodnar2016direct}
Bodnar, T., A.~K. Gupta, and N.~Parolya (2016).
\newblock Direct shrinkage estimation of large dimensional precision matrix.
\newblock {\em Journal of Multivariate Analysis\/}~{\em 146}, 223--236.

\bibitem[\protect\citeauthoryear{Bodnar, Okhrin, and Parolya}{Bodnar
  et~al.}{2019}]{bodnar2019optimal}
Bodnar, T., O.~Okhrin, and N.~Parolya (2019).
\newblock Optimal shrinkage estimator for high-dimensional mean vector.
\newblock {\em Journal of Multivariate Analysis\/}~{\em 170}, 63--79.

\bibitem[\protect\citeauthoryear{Boyd, Parikh, Chu, Peleato, Eckstein,
  et~al.}{Boyd et~al.}{2011}]{boydADMM:2011}
Boyd, S., N.~Parikh, E.~Chu, B.~Peleato, J.~Eckstein, et~al. (2011).
\newblock Distributed optimization and statistical learning via the alternating
  direction method of multipliers.
\newblock {\em Foundations and Trends{\textregistered} in Machine
  learning\/}~{\em 3\/}(1), 1--122.

\bibitem[\protect\citeauthoryear{Bryzgalova, Lerner, Lettau, and
  Pelger}{Bryzgalova et~al.}{2022}]{bryzgalova2022missing}
Bryzgalova, S., S.~Lerner, M.~Lettau, and M.~Pelger (2022).
\newblock Missing financial data.
\newblock {\em Available at SSRN 4106794\/}.

\bibitem[\protect\citeauthoryear{Bryzgalova, Pelger, and Zhu}{Bryzgalova
  et~al.}{2020}]{bryzgalova2020forest}
Bryzgalova, S., M.~Pelger, and J.~Zhu (2020).
\newblock Forest through the trees: Building cross-sections of stock returns.
\newblock {\em Available at SSRN 3493458\/}.

\bibitem[\protect\citeauthoryear{Carhart}{Carhart}{1997}]{carhart1997persistence}
Carhart, M.~M. (1997).
\newblock On persistence in mutual fund performance.
\newblock {\em The Journal of finance\/}~{\em 52\/}(1), 57--82.

\bibitem[\protect\citeauthoryear{Chitsiripanich, Paolella, Polak, and
  Walker}{Chitsiripanich et~al.}{2022}]{FracMom:22}
Chitsiripanich, S., M.~S. Paolella, P.~Polak, and P.~S. Walker (2022).
\newblock Momentum without crashes.
\newblock {\em Swiss Finance Institute Research Paper\/}~(22-87).

\bibitem[\protect\citeauthoryear{DeMiguel, Garlappi, Nogales, and
  Uppal}{DeMiguel et~al.}{2009}]{demiguel2009generalized}
DeMiguel, V., L.~Garlappi, F.~J. Nogales, and R.~Uppal (2009).
\newblock A generalized approach to portfolio optimization: Improving
  performance by constraining portfolio norms.
\newblock {\em Management science\/}~{\em 55\/}(5), 798--812.

\bibitem[\protect\citeauthoryear{DeMiguel, Garlappi, and Uppal}{DeMiguel
  et~al.}{2009}]{DGU:09}
DeMiguel, V., L.~Garlappi, and R.~Uppal (2009).
\newblock Optimal versus naive diversification: How inefficient is the $1/n$
  portfolio strategy?
\newblock {\em Review of Financial Studies\/}~{\em 22\/}(5), 1915--1953.

\bibitem[\protect\citeauthoryear{Fama and French}{Fama and
  French}{1993}]{FAMA1993}
Fama, E.~F. and K.~R. French (1993).
\newblock Common risk factors in the returns on stocks and bonds.
\newblock {\em Journal of financial economics\/}~{\em 33\/}(1), 3--56.

\bibitem[\protect\citeauthoryear{Fama and French}{Fama and
  French}{2015}]{fama2015five}
Fama, E.~F. and K.~R. French (2015).
\newblock A five-factor asset pricing model.
\newblock {\em Journal of Financial Economics\/}~{\em 116\/}(1), 1--22.

\bibitem[\protect\citeauthoryear{Fan, Liao, and Mincheva}{Fan
  et~al.}{2013}]{fan2013large}
Fan, J., Y.~Liao, and M.~Mincheva (2013).
\newblock Large covariance estimation by thresholding principal orthogonal
  complements.
\newblock {\em Journal of the Royal Statistical Society Series B: Statistical
  Methodology\/}~{\em 75\/}(4), 603--680.

\bibitem[\protect\citeauthoryear{Feng, Giglio, and Xiu}{Feng
  et~al.}{2020}]{feng2020taming}
Feng, G., S.~Giglio, and D.~Xiu (2020).
\newblock Taming the factor zoo: A test of new factors.
\newblock {\em The Journal of Finance\/}~{\em 75\/}(3), 1327--1370.

\bibitem[\protect\citeauthoryear{Freyberger, Neuhierl, and Weber}{Freyberger
  et~al.}{2020}]{freyberger2020dissecting}
Freyberger, J., A.~Neuhierl, and M.~Weber (2020).
\newblock Dissecting characteristics nonparametrically.
\newblock {\em The Review of Financial Studies\/}~{\em 33\/}(5), 2326--2377.

\bibitem[\protect\citeauthoryear{Goyal and Saretto}{Goyal and
  Saretto}{2022}]{goyal2022equity}
Goyal, A. and A.~Saretto (2022).
\newblock Are equity option returns abnormal? ipca says no.
\newblock {\em IPCA Says No (August 19, 2022)\/}.

\bibitem[\protect\citeauthoryear{Hastie, Tibshirani, and Wainwright}{Hastie
  et~al.}{2015}]{Hastie:2015}
Hastie, T., R.~Tibshirani, and M.~Wainwright (2015).
\newblock Statistical learning with sparsity.
\newblock {\em Monographs on statistics and applied probability\/}~{\em 143},
  143.

\bibitem[\protect\citeauthoryear{Hediger, N{\"a}f, Paolella, and Polak}{Hediger
  et~al.}{2021}]{hediger2021heterogeneous}
Hediger, S., J.~N{\"a}f, M.~S. Paolella, and P.~Polak (2021).
\newblock Heterogeneous tail generalized common factor modeling.
\newblock {\em Swiss Finance Institute Research Paper\/}~(21-73).

\bibitem[\protect\citeauthoryear{Jegadeesh and Titman}{Jegadeesh and
  Titman}{1993}]{jegadeesh1993mom}
Jegadeesh, N. and S.~Titman (1993).
\newblock Returns to buying winners and selling losers: Implications for stock
  market efficiency.
\newblock {\em The Journal of Finance\/}~{\em 48\/}(1), 65--91.

\bibitem[\protect\citeauthoryear{Kelly, Pruitt, and Su}{Kelly
  et~al.}{2019}]{KELLY2019501}
Kelly, B.~T., S.~Pruitt, and Y.~Su (2019).
\newblock Characteristics are covariances: A unified model of risk and return.
\newblock {\em Journal of Financial Economics\/}~{\em 134\/}(3), 501--524.

\bibitem[\protect\citeauthoryear{Kourtis, Dotsis, and Markellos}{Kourtis
  et~al.}{2012}]{kourtis2012parameter}
Kourtis, A., G.~Dotsis, and R.~N. Markellos (2012).
\newblock Parameter uncertainty in portfolio selection: Shrinking the inverse
  covariance matrix.
\newblock {\em Journal of Banking \& Finance\/}~{\em 36\/}(9), 2522--2531.

\bibitem[\protect\citeauthoryear{Ledoit and Wolf}{Ledoit and
  Wolf}{2004}]{Ledoit:04}
Ledoit, O. and M.~Wolf (2004).
\newblock Honey, i shrunk the sample covariance matrix.
\newblock {\em The Journal of Portfolio Management\/}~{\em 30\/}(4), 110--119.

\bibitem[\protect\citeauthoryear{Ledoit and Wolf}{Ledoit and
  Wolf}{2012}]{Ledoit:12}
Ledoit, O. and M.~Wolf (2012).
\newblock Nonlinear shrinkage estimation of large-dimensional covariance
  matrices.
\newblock {\em The Annals of Statistics\/}~{\em 40\/}(2), 1024 -- 1060.

\bibitem[\protect\citeauthoryear{Ledoit and Wolf}{Ledoit and
  Wolf}{2020a}]{ledoit2020analytical}
Ledoit, O. and M.~Wolf (2020a).
\newblock Analytical nonlinear shrinkage of large-dimensional covariance
  matrices.
\newblock {\em The Annals of Statistics\/}~{\em 48\/}(5), 3043--3065.

\bibitem[\protect\citeauthoryear{Ledoit and Wolf}{Ledoit and
  Wolf}{2020b}]{LedoitWolf:20}
Ledoit, O. and M.~Wolf (2020b, 06).
\newblock {The Power of (Non-)Linear Shrinking: A Review and Guide to
  Covariance Matrix Estimation}.
\newblock {\em Journal of Financial Econometrics\/}~{\em 20\/}(1), 187--218.

\bibitem[\protect\citeauthoryear{Ledoit and Wolf}{Ledoit and
  Wolf}{2022}]{LedoitWolf:22}
Ledoit, O. and M.~Wolf (2022).
\newblock Quadratic shrinkage for large covariance matrices.
\newblock {\em Bernoulli\/}~{\em 28\/}(3).

\bibitem[\protect\citeauthoryear{Lettau and Pelger}{Lettau and
  Pelger}{2020}]{lettau2020factors}
Lettau, M. and M.~Pelger (2020).
\newblock Factors that fit the time series and cross-section of stock returns.
\newblock {\em The Review of Financial Studies\/}~{\em 33\/}(5), 2274--2325.

\bibitem[\protect\citeauthoryear{Li}{Li}{2015}]{li2015sparse}
Li, J. (2015).
\newblock Sparse and stable portfolio selection with parameter uncertainty.
\newblock {\em Journal of Business \& Economic Statistics\/}~{\em 33\/}(3),
  381--392.

\bibitem[\protect\citeauthoryear{Lintner}{Lintner}{1965}]{lintner1965security}
Lintner, J. (1965).
\newblock Security prices, risk, and maximal gains from diversification.
\newblock {\em The Journal of Finance\/}~{\em 20\/}(4), 587--615.

\bibitem[\protect\citeauthoryear{Lunde, Shephard, and Sheppard}{Lunde
  et~al.}{2016}]{lunde2016econometric}
Lunde, A., N.~Shephard, and K.~Sheppard (2016).
\newblock Econometric analysis of vast covariance matrices using composite
  realized kernels and their application to portfolio choice.
\newblock {\em Journal of Business \& Economic Statistics\/}~{\em 34\/}(4),
  504--518.

\bibitem[\protect\citeauthoryear{Markowitz}{Markowitz}{1952}]{Ma52}
Markowitz, H. (1952).
\newblock Modern portfolio theory.
\newblock {\em Journal of Finance\/}~{\em 7\/}(11), 77--91.

\bibitem[\protect\citeauthoryear{Mossin}{Mossin}{1966}]{mossin1966equilibrium}
Mossin, J. (1966).
\newblock Equilibrium in a capital asset market.
\newblock {\em Econometrica: Journal of the econometric society\/}, 768--783.

\bibitem[\protect\citeauthoryear{Paolella and Polak}{Paolella and
  Polak}{2015}]{PaPo:15c}
Paolella, M.~S. and P.~Polak (2015).
\newblock Portfolio selection with active risk monitoring.
\newblock {\em Swiss Finance Institute Research Paper\/}~(15-17).

\bibitem[\protect\citeauthoryear{Paolella, Polak, Polino, and Walker}{Paolella
  et~al.}{2022}]{PaPoPoWa:19}
Paolella, M.~S., P.~Polak, A.~Polino, and P.~S. Walker (2022).
\newblock Risk parity versus risk minimization portfolio allocation under
  heavy-tailed returns.
\newblock Working Paper.

\bibitem[\protect\citeauthoryear{Paolella, Polak, and Walker}{Paolella
  et~al.}{2019}]{paolella2019regime}
Paolella, M.~S., P.~Polak, and P.~S. Walker (2019).
\newblock Regime switching dynamic correlations for asymmetric and fat-tailed
  conditional returns.
\newblock {\em Journal of Econometrics\/}~{\em 213\/}(2), 493--515.

\bibitem[\protect\citeauthoryear{Paolella, Polak, and Walker}{Paolella
  et~al.}{2021}]{Paolella:2021}
Paolella, M.~S., P.~Polak, and P.~S. Walker (2021).
\newblock A non-elliptical orthogonal garch model for portfolio selection under
  transaction costs.
\newblock {\em Journal of Banking \& Finance\/}~{\em 125}, 106046.

\bibitem[\protect\citeauthoryear{Pedersen}{Pedersen}{2015}]{Pedersen:15}
Pedersen, L.~H. (2015).
\newblock {\em Efficiently Inefficient: How Smart Money Invests and Market
  Prices Are Determined}.
\newblock Prienceton University Press.

\bibitem[\protect\citeauthoryear{Roncalli}{Roncalli}{2013}]{roncalli_RP:13}
Roncalli, T. (2013).
\newblock {\em Introduction to risk parity and budgeting}.
\newblock CRC Press.

\bibitem[\protect\citeauthoryear{Sharpe}{Sharpe}{1964}]{sharpe1964capital}
Sharpe, W.~F. (1964).
\newblock Capital asset prices: A theory of market equilibrium under conditions
  of risk.
\newblock {\em The Journal of Finance\/}~{\em 19\/}(3), 425--442.

\bibitem[\protect\citeauthoryear{Stellato, Banjac, Goulart, Bemporad, and
  Boyd}{Stellato et~al.}{2020}]{osqp}
Stellato, B., G.~Banjac, P.~Goulart, A.~Bemporad, and S.~Boyd (2020).
\newblock {OSQP}: an operator splitting solver for quadratic programs.
\newblock {\em Mathematical Programming Computation\/}~{\em 12\/}(4), 637--672.

\bibitem[\protect\citeauthoryear{Stock and Watson}{Stock and
  Watson}{2002}]{stock2002forecasting}
Stock, J.~H. and M.~W. Watson (2002).
\newblock Forecasting using principal components from a large number of
  predictors.
\newblock {\em Journal of the American statistical association\/}~{\em
  97\/}(460), 1167--1179.

\bibitem[\protect\citeauthoryear{Tibshirani}{Tibshirani}{1996}]{tibshirani1996}
Tibshirani, R. (1996).
\newblock Regression shrinkage and selection via the lasso.
\newblock {\em Journal of the Royal Statistical Society: Series B
  (Methodological)\/}~{\em 58\/}(1), 267--288.

\bibitem[\protect\citeauthoryear{Treynor}{Treynor}{1961}]{treynor1961market}
Treynor, J.~L. (1961).
\newblock Market value, time, and risk.
\newblock {\em Time, and Risk (August 8, 1961)\/}.

\bibitem[\protect\citeauthoryear{Tsai and Tsay}{Tsai and
  Tsay}{2010}]{tsai2010constrained}
Tsai, H. and R.~S. Tsay (2010).
\newblock Constrained factor models.
\newblock {\em Journal of the American Statistical Association\/}~{\em
  105\/}(492), 1593--1605.

\bibitem[\protect\citeauthoryear{Wang, Pan, Tong, and Zhu}{Wang
  et~al.}{2015}]{wang2015shrinkage}
Wang, C., G.~Pan, T.~Tong, and L.~Zhu (2015).
\newblock Shrinkage estimation of large dimensional precision matrix using
  random matrix theory.
\newblock {\em Statistica Sinica\/}, 993--1008.

\bibitem[\protect\citeauthoryear{Wang, Tong, Cao, and Miao}{Wang
  et~al.}{2014}]{wang2014non}
Wang, C., T.~Tong, L.~Cao, and B.~Miao (2014).
\newblock Non-parametric shrinkage mean estimation for quadratic loss functions
  with unknown covariance matrices.
\newblock {\em Journal of Multivariate Analysis\/}~{\em 125}, 222--232.

\end{thebibliography}

\newpage

\section*{Appendix: Description of Asset Specific Factors}

In Table \ref{table:characteristics}, we list the details of the firm specific characteristics used in the factors models.

{
\captionsetup{labelfont={color=black},textfont={color=black}}
\begin{table}[h]
\centering
\caption{\protect \footnotesize Acronyms and Factor Names}
\label{table:characteristics}
\scriptsize % Applying the scriptsize command
\begin{tabularx}{\textwidth}{ll|ll}
\toprule
\textbf{Acronym} & \textbf{Characteristic Name} & \textbf{Acronym} & \textbf{Characteristic Name} \\
\midrule
A2ME & Assets-to-market cap & Lev & Leverage \\
AT & Total assets & LME$^*$ & Size \\
ATO & Net sales over lagged net operating assets & LTurnover$^*$ & Turnover \\
BEME$^*$ & Ratio of book value of equity to market value of equity & NOA & Net operating assets \\
Beta$^*$ & CAPM beta & OA & Operating accruals \\
C & Ratio of cash and short-term investments to total assets & OL & Operating leverage \\
CTO & Capital turnover & PCM & Price-to-cost margin \\
DTO & Daily turnover & PM & Profit margin \\
Lt\_Rev$^*$ & 5 Years Long-term reversal & Rel to High & Closeness to 52-week high \\
E2P & Earnings to price & Q & Tobin's Q \\
AC$^*$ & Change in operating working capital & RNA & Return on net operating assets \\
Idio vol$^*$ & Idiosyncratic volatility & ROA & Return-on-assets \\
Mktcap$^*$ & Market capitalization & ROE & Return-on-equity \\
r12-2$^*$ & Momentum & r12-7 & Intermediate momentum \\
r2-1$^*$ & Short-term reversal & r36-13 & 3 Years Long-term reversal \\
S2P & Sales-to-price & SGA2S & SG\&A to sales \\
SUV & Standard unexplained volume &  &  \\
\bottomrule
\end{tabularx}
\textit{Note:} $^*$ denotes the characteristics used in AP-Trees managed portfolios.
\end{table}}

\end{document}